\documentclass[twocolumn]{article}
\usepackage[utf8]{inputenc}
\usepackage[margin=17mm]{geometry}
\usepackage{fancyhdr}
\usepackage{amsmath}
\usepackage{amsfonts}
\usepackage{amssymb}
\usepackage{graphicx}
\usepackage{verbatim}

\usepackage[long,nodayofweek]{datetime}

\usepackage{authblk}
\usepackage{hyperref}

\providecommand{\keywords}[1]
{
  \small
  \textbf{\textit{Keywords---}} #1
}
\title{Impact of non-tidal station loading in LLR}
\author[1,2]{Vishwa Vijay Singh\footnote{Corresponding author:\newline Vishwa Vijay Singh, email: \href{mailto:singh@ife.uni-hannover.de}{singh@ife.uni-hannover.de}}}
\author[1]{Liliane Biskupek}
\author[1]{Jürgen Müller}
\author[1,3,4]{Mingyue Zhang}
\affil[1]{\small Institute of Geodesy (IfE),\ Leibniz University Hannover,\ Schneiderberg 50,\ 30167 Hannover,\ Germany}
\affil[2]{\small Institute for Satellite Geodesy and Inertial Sensing, German Aerospace Center (DLR), Callinstraße 36, 30167 Hannover, Germany}
\affil[3]{\small State Key Laboratory of Geodesy and Earth's Dynamics, Institute of Geodesy and Geophysics, APM, Chinese Academy of Sciences, Wuhan 430077, China}
\affil[4]{\small College of Earth and Planetary Sciences, University of Chinese Academy of Sciences, Beijing 100049, China}

\usepackage{framed}
\usepackage{csquotes}
\usepackage{enumitem}
\usepackage[para,online,flushleft]{threeparttable}
\usepackage{breakcites}
\usepackage{multirow}
\usepackage{natbib}
\usepackage{url}

\begin{document}

\maketitle
\thispagestyle{fancy}





\begin{abstract}
Lunar Laser Ranging (LLR) measures the distance between observatories on Earth and retro-reflectors on Moon since 1970. In this paper, we study the effect of non-tidal station loading (NTSL) in the analysis of LLR data. We add the non-tidal loading effect provided by three data centres: the German Research Centre for Geosciences (GFZ), the International Mass Loading Service (IMLS) and EOST loading service of University of Strasbourg in France, as observation level corrections of the LLR observatories in our analysis. This effect causes deformations of the Earth surface up to the centimetre level. Its addition in the Institute of Geodesy (IfE) LLR model, it leads to a change in the uncertainties (3-$\sigma$ values) of the station coordinates resulting in up to a 0.60\% improvement, an improvement in the post-fit LLR residuals of up to 9\%, and a decrease in the power of the annual signal in the LLR post-fit residuals of up to 57\%.
\end{abstract}



\section{Introduction}
Lunar laser ranging (LLR) is the measurement of round trip travel times of short laser pulses between observatories on the Earth and retro-reflectors on the Moon. There are five retro-reflectors on the Moon, and measurements have primarily been carried out from six observatories on Earth, details of which are given in Table \ref{tab:ife_res}. For each returning laser pulse, the round-trip travel time (Earth-Moon-Earth) is observed. As the amount of signal loss of the laser pulse is enormous, it is necessary to collect measurements for 1 to 15 minutes. Of these, a statistically secured mean value is computed, a so called normal point (NP). Details of the LLR measurement process can be found, for example, in \citet{murp13} and \citet{mueller19}. The NP is treated as the actual observable of LLR.  The recent NP are provided by the CDDIS and be downloaded from the website\footnote{\url{https://cddis.nasa.gov/Data_and_Derived_Products/SLR/Lunar_laser_ranging_data.html}}. The LLR data is available from 1969, and at present, the Institute of Geodesy (IfE) LLR dataset contains 26839 NPs, and spans from 1970 - 2019.

\begin{table}[!ht]\centering
\caption{Details of LLR observatories and their observations used within IfE normal point (NP) file.}\label{tab:ife_res}
\begin{tabular}{p{2.5cm}cccc}
\hline\noalign{\smallskip}
Name &Abbreviation &NPs &Timespan \\
\noalign{\smallskip}\hline\noalign{\smallskip}
{Observatoire de la Côte d’Azur, Grasse, France} &\multirow{1}[7]{*}{OCA} &\multirow{1}[7]{*}{16298} &\multirow{1}[7]{*}{\shortstack{1984 - 2005, \\ 2009 - 2019}} \\
\noalign{\smallskip}\hline\noalign{\smallskip}
{Matera Laser Ranging Observatory, Matera, Italy} &\multirow{1}[11]{*}{MLRO} &\multirow{1}[11]{*}{240} &\multirow{1}[11]{*}{\shortstack{2003 - 2004, \\ 2010 - 2019}} \\
\noalign{\smallskip}\hline\noalign{\smallskip}
{Wettzell Laser Ranging System, Wettzell, Germany} &\multirow{1}[11]{*}{WLRS} &\multirow{1}[11]{*}{48} &\multirow{1}[11]{*}{\shortstack{1994, 1996, \\ 2018 - 2019}} \\
\noalign{\smallskip}\hline\noalign{\smallskip}
{Apache Point Observatory Lunar Laser-ranging Operation, in New Mexico, USA} &\multirow{1}[13]{*}{APOLLO} &\multirow{1}[13]{*}{2585} &\multirow{1}[13]{*}{2006 - 2016} \\
\noalign{\smallskip}\hline\noalign{\smallskip}
{Lunar Laser Ranging Experiment Observatory, Hawaii, USA} &\multirow{1}[11]{*}{LURE} &\multirow{1}[11]{*}{755} &\multirow{1}[11]{*}{1984 - 1990} \\
\noalign{\smallskip}\hline\noalign{\smallskip}
\multirow{3}[4]{2.5cm}{McDonald Laser Ranging Station, Texas, USA} &McDonald &3070 &1970 - 1985 \\
\noalign{\smallskip}\cline{2-4}\noalign{\smallskip}&MLRS1 &710 &1983 - 1988 \\
\noalign{\smallskip}\cline{2-4}\noalign{\smallskip}&MLRS2 &3133 &1988 - 2013 \\
\noalign{\smallskip}\hline
\end{tabular}
\end{table}

As LLR has the longest observation time series of all space geodetic techniques \citep{mueller19}, it allows the determination of a variety of parameters of the Earth–Moon dynamics, for example, the mass of the Earth–Moon system, the lunar orbit and libration parameters \citep{de430_other, pav16}; and it leads to improvements in the solar system ephemerides \citep{jplde}, terrestrial and celestial reference frames and coordinates of observatories and reflectors \citep{mueller2009,hof_etal18}, selenophysics \citep{murp13,hofmann17,viswa19}, and gravitational physics, i.e. tests of Einstein's relativity theory, for example, strong equivalence principle, metric or preferred-frame effects, variation of the gravitational constant \citep{will06,mueller_etal12,hof_mu18,Zhang2020}. LLR can also be used to provide tests of Earth orientation parameters \citep{bisk2015,hof_etal18}.

The Earth's crust is continuously deforming, due to which the positions of the observatories on Earth change over time. Various geophysical process contribute to the deformation of the crust, which can be estimated by different models. The 2010 conventions of the International Earth Rotation and Reference Systems Service (IERS) \citep{Petit2010} provide details of the models recommended to be used for instantaneous calculation of deformations of reference points on the crust.

The deformations of the crust due to redistribution of masses in atmospheric, ocean, and land water mass has both tidal and non-tidal loading (NTL) components. The IERS 2010 conventions do not recommend the addition of NTL deformations in the calculation of the displacement of a reference points. The IERS, however, established the Global Geophysical Fluids Center (GGFC)  in 1998, which has different bureaus responsible for research and data provision related to the redistribution of masses in atmosphere, oceans, and hydrological (land water) systems. These bureaus, amongst other products, provide time series of NTLs over different time spans, based on calculations using numerical weather models and Green's functions \citep{farrell72,gfz_ntsl,imls_ntsl}, which can be added as observation level corrections in the calculation of the instantaneous position of a reference site.

NTL plays a special role in optical observation techniques (satellite laser ranging (SLR) and LLR) as their observations can only be performed during clear sky conditions, creating a difference in their results in comparison with microwave observation techniques (such as Very Long Baseline Interferometry (VLBI) and Global Navigation Satellite Systems (GNSS)) \citep{otsubo_04,sonsica_13,bury_2019}. This weather-dependent-effect on the results is called the Blue-Sky effect. The accuracy of the loading effect due to NTL, as pointed out by \cite{glomsda_etal2020}, has improved over the past years due to the improved accuracy of the numerical weather models used for its calculation \citep{mpiom,merra2,era5,gfz_ntsl}, and therefore addition of NTL can be beneficial in geodetic analyses.

The effect of NTL has already been studied in VLBI (\citep{schuh2003,glomsda_etal2020}, and others), GNSS (\citep{boy2008_nto,dach_etal2010,vanDam12,nordman_etal2015,memim20}, and others), and SLR (\citep{sonsica_13,bury_2019}, and others). Their results have mentioned that addition of displacements due to NTL leads to an improvement, most significantly in reduction of seasonal signals.

In this paper, Section \ref{sec:ntsl} describes the NTL and its components and compares the data from different data centres for all loading components. Section \ref{sec:NTL_LUNAR} contains the results of implementing all loadings of NTL from different data centres in LLR analysis. Section \ref{sec:conclusions} gives the conclusions and addresses further aspects in this context.

\section{Non-tidal loading datasets}\label{sec:ntsl}



As mentioned in section 1, the redistribution of masses in atmosphere, ocean, and land water causes displacements which have NTL components. These displacements for any particular point $X$ are calculated based on \cite{farrell72}, which converts pressure differences from a mean pressure value to horizontal and vertical displacement components. The calculation involves integration over an area $A$ around the point $X$, weighting the pressure at other points $X'$ within the area using a Green's function. The Green's function is dependent on the angular distance between the point $X$ and $X'$, given as:
\begin{equation}
    Gr(\cos{\beta}) = \frac{G R}{g}\sum_{n=1}^{\infty}h_{n}'P(\cos{\beta}),
\end{equation}
where $\beta$ is the angular distance between point $X$ and $X'$, $h_{n}'$ is the load Love number, $P(\cos{\beta})$ the Legendre polynomial, $G$ the gravitational constant, $g$ the mean surface gravity, and $R$ the mean radius of the Earth.

The deformation (here, up component) at point $X$ is then calculated as:
\begin{equation}\label{eq:disp}
    \zeta_{up}(X,t) = \int_{A}\frac{\Delta p}{g}Gr(\cos{\beta})dA,
\end{equation}
where $\Delta p$ is the pressure difference at the point of integration, $dA$ is surface element defined by $X$ and $X'$. The procedure for displacement calculation due to NTL is described in detail by \cite{petrov_boy04,gfz_ntsl,imls_ntsl} and others.

The pressure values are obtained from various different numerical weather models (NWMs), which consider different effects for their own calculation.

According to the GGFC website\footnote{\url{http://loading.u-strasbg.fr/GGFC/}}, the NTL data is provided by the following official centres:
\begin{enumerate}[itemsep=-1ex]
    \item EOST loading service, Univerity of Strasbourg, France,
    \item GGOS Atmosphere at Vienna (VMF), Technical University Vienna, Austria,
    \item German Research Centre for Geosciences (GFZ), Potsdam, Germany, and
    \item University of Luxembourg (ULux), Luxembourg.
\end{enumerate}
Additionally, NTL data can also be obtained from the International Mass Loading Service (IMLS)\footnote{\url{http://massloading.net/}}. Not all data centres provide all three non-tidal loadings (atmosphere: NTAL, ocean: NTOL, and hydrological: HYDL), and some of the data centres provide loadings calculated from more than one NWM. In Table \ref{tab:model_list}, we give a list of the loadings and their corresponding NWMs considered in this study.

\begin{table}[!ht]
\caption{Details of the loading components and their corresponding numerical weather models (NWM) of different data centres.}
\begin{threeparttable}
  \centering
    \begin{tabular}{p{0.85cm}p{1.10cm}p{1.75cm}p{1.35cm}p{1.25cm}}
    \hline\noalign{\smallskip}
    Dataset & Timespan  & Loading Component & NWM & Temporal Resolution \\
    \noalign{\smallskip}\hline\noalign{\smallskip}
    \multirow{4}{0.85cm}{GFZ} & \multirow{3}{1cm}{1976 - present} & Atmospheric & ECMWF & 3h or 24h \\
      &   & Oceanic & MPIOM & 3h or 24h \\
      &   & Hydrological & LSDM & 24h \\
      &   & Sea level & LSDM + ECMWF & 24h \\
    \noalign{\smallskip}\hline\noalign{\smallskip}
    \multirow{3}{0.85cm}{IMLS} & \multirow{3}{1cm}{1980 - present} & \rule{0pt}{2.5ex}Atmospheric & MERRA2 & 6h \\
      &   & Oceanic & MPIOM & 3h \\
      &   & Hydrological & MERRA2 & 3h \\
    \noalign{\smallskip}\hline\noalign{\smallskip}
    \multirow{3}{0.85cm}{EOST} & \multirow{2}{1cm}{1980 - present} & \rule{0pt}{2.5ex}Atmospheric & MERRA2 & 1h \\
      &   & Oceanic\tnote{1} & ECCO2 & 24h \\
      &   & Hydrological & MERRA2 & 1h \\
    \noalign{\smallskip}\hline\noalign{\smallskip}
    {VMF} & {1994 - present} & \rule{0pt}{2.5ex}Atmospheric & ECWMF & 6h \\
    \noalign{\smallskip}\hline\noalign{\smallskip}
    {ULux} & {1980 - 2015} & \rule{0pt}{2.5ex}Atmospheric & NCEP & 6h \\
    \noalign{\smallskip}\hline
    \end{tabular}%
    \begin{tablenotes}
     \item[*] starting 1992\\
     \item ECMWF: European Centre for Medium-Range Weather Forecasts\\
     \item MPIOM: Max-Planck-Institute Global Ocean/Sea-Ice Model\\
     \item LSDM: Land Surface Discharge Model\\
     \item MERRA2: Modern-Era Retrospective analysis for Research and Applications, Version 2\\
     \item NCEP: National Center for Environmental Protection (USA)\\
     \item ECCO2: Estimating the Circulation and Climate of the Ocean, version 2\\
   \end{tablenotes}
    \end{threeparttable}%
  \label{tab:model_list}
\end{table}%

For IMLS and EOST dataset, in addition to the the loadings from the NWMs mentioned in Table \ref{tab:model_list}, other options are also available. However, the other NWMs have limitations, such as a shorter time span, discontinued NWM, et cetera, and therefore were not used for this study.

As it can be seen from Table \ref{tab:model_list}, the datasets from different centres have some differences in the temporal resolution and the models used for their computation. Slight differences in the displacements could also occur due to the procedure used to compute the integral to get them. All datasets use Green’s functions to compute the NTL displacements, however, the GFZ and IMLS mention two special approaches to solve the convolution integrals (like Eq \ref{eq:disp}). The GFZ uses a patched version to be able to reduce the computation time to obtain the displacements \citep{gfz_ntsl}. It applies a high spatial resolution for nearby pressure fields and a lower spatial resolution for those far away, which are combined used fast interpolation techniques. The IMLS uses a spherical harmonic transformation approach to solve the integral. The algorithm for this transformation is described in \citep{imls_ntsl}.



\subsection{Non-tidal atmospheric loading}\label{subsec:ntal}
The effect of NTAL has been extensively studied within different space geodetic techniques, such as \cite{vanDam_etal_1994,tregoning_2009,tregoning_2011,sonsica_13,koenig_etal_2018,bury_2019} and others. The atmospheric pressure loading (APL) over the Earth has a tidal and non-tidal component. Both components of APL are modelled separately, and can cause up to cm level deformations of the Earth surface \citep{petrov_boy04}.


The application of NTAL commonly assumes the response of ocean to the atmospheric loading to be that of an inverted barometer (IB), i.e. an increase in atmospheric pressure at any point of the ocean is directly compensated by an equivalent decrease in sea level at this point \cite{petrov_boy04,glomsda_etal2020}. This also leads to a decrease in ocean bottom pressure of equal size. Height variations of the ocean for periods longer than 5–20 days are successfully represented by this model, however, for shorter periods this model does not capture the height variations correctly \citep{wunsch_stammer_97}. According to \cite{vanDam_Wahr_1987}, points on Earth within a few hundred kilometres of the coast are affected by the choice of the oceanic response (IB or non-inverted barometer (NIB)) to APL, showing a higher value of deformation for NIB than the deformation at the same point calculated by an IB model. Inland observation points are not significantly affected by the choice of the ocean response model.

\begin{figure*}[!ht]
	\centering
	\includegraphics[width=0.74\textwidth]{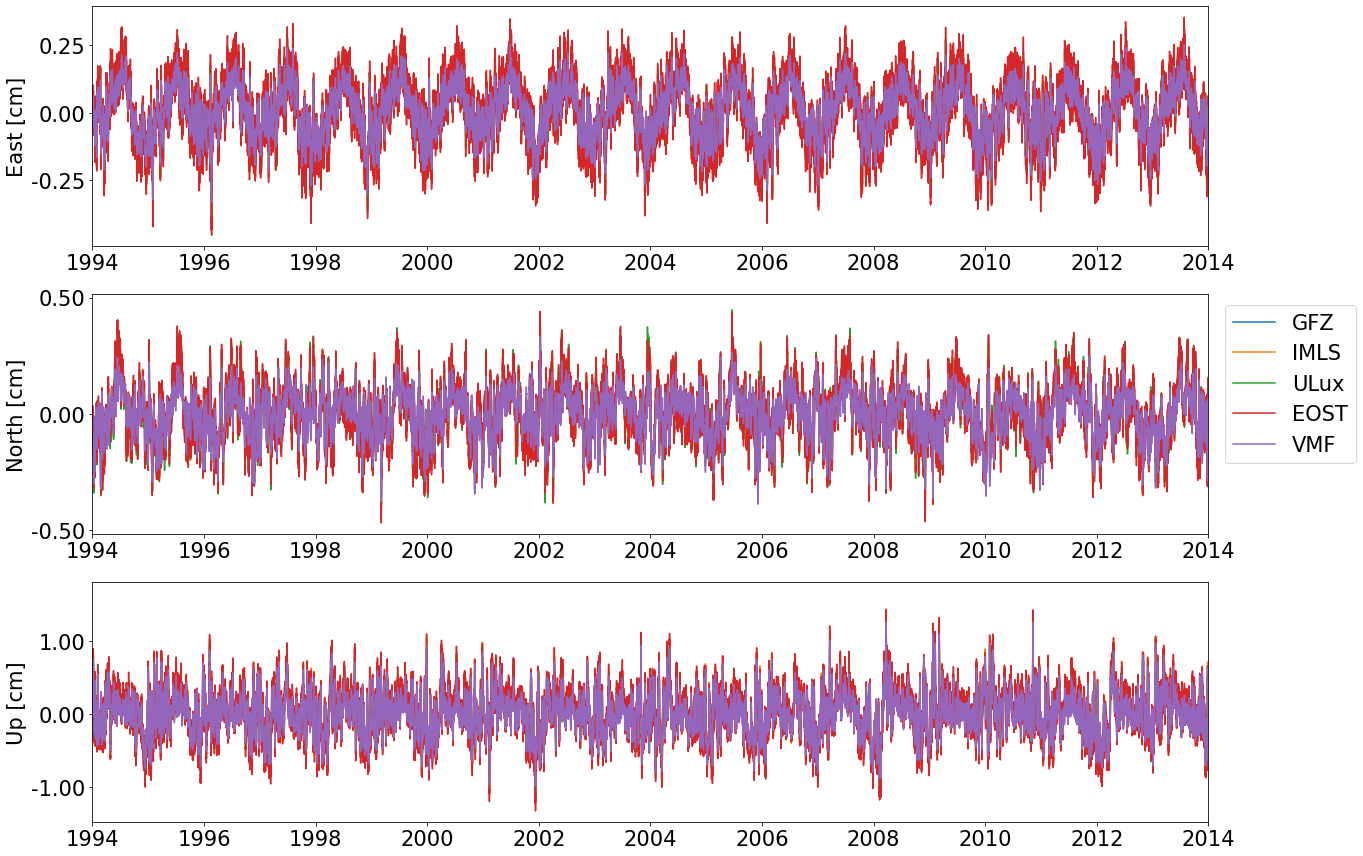}
	\caption{Effect of NTAL at the OCA station from 1994 to 2014 for GFZ, IMLS, ULux, EOST, and VMF datasets.}\label{fig:OCA_NTAL}
\end{figure*}

From the Figure \ref{fig:OCA_NTAL}, it can be seen that the NTAL has the highest effect in the up component of the displacement, ranging between $\pm$1 cm. The horizontal displacement due to NTAL lies between $\pm$0.45 cm (North) and between $\pm$0.3 cm (East). The effect of NTAL has the highest contribution of all NTL effects for inland observatories. For NTAL, all input time series almost completely overlap each other for all (Up, North, East) components, with the exception of VMF, which shows differences in the horizontal components of loading for some stations compared to other datasets. The maximum of these differences, for LLR observatories, are observed for the APOLLO and McDonald stations (not shown), where the differences in the range of the horizontal components of VMF compared to the GFZ dataset are up to 45\% for both stations. These differences could be due to different land-sea masks, resolution, weather models, computation method, et cetera. However, as the horizontal components of NTAL are less significant than the vertical components, and as the effect of horizontal components of NTAL is only up to a few millimetres, it is not expected to produce significant differences in the time series of geodetic observations between results obtained from VMF and other NTAL datasets.


\subsection{Non-tidal oceanic loading}\label{subsec:ntol}
The ocean water redistribution by atmospheric circulation, inflow and outflow of ocean water, and changes in the total atmospheric mass over the oceans primarily cause NTOL deformations \citep{glomsda_etal2020}. It plays an important role in different space geodetic techniques, and its effect has been studied, for example, by \cite{williams_and_penna,vanDam12,memin_etal_2014,orerio_etal_2018,memim20}, and others.

\begin{figure*}[!ht]
	\centering
	\includegraphics[width=0.74\textwidth]{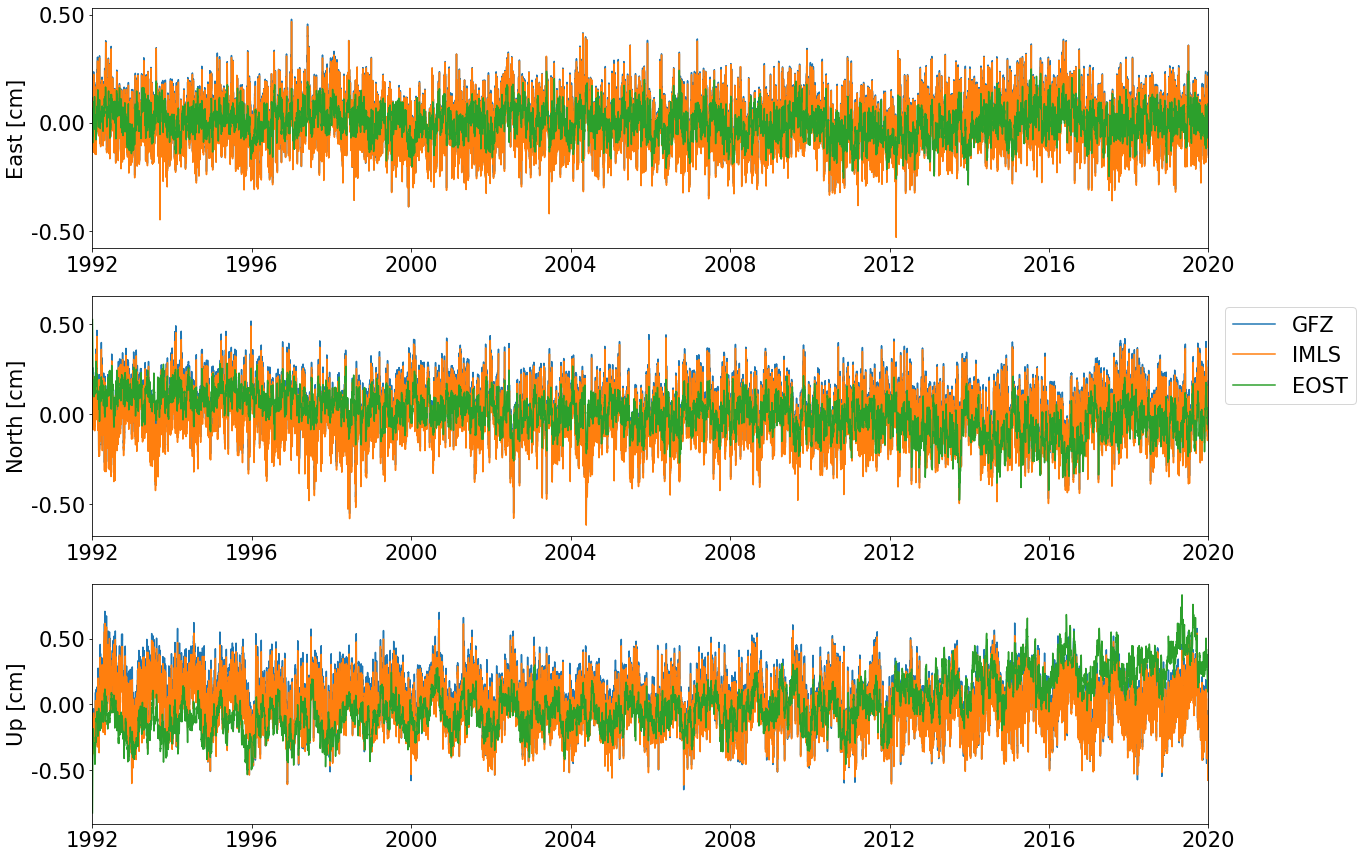}
	\caption{Effect of NTOL at the OCA station from 1992 to 2020 for GFZ, IMLS, and EOST datasets.}\label{fig:OCA_NTOL}
\end{figure*}

Figures \ref{fig:OCA_NTOL} shows the NTOL at the OCA station. All LLR stations show a similar trend for NTOL, i.e. GFZ and IMLS NTOL time series are similar, and the EOST time series differs, but stays within the same range as GFZ and IMLS, $\pm$0.65 cm for Up component, $\pm$0.50 cm for North component, and -0.50 cm to 0.40 cm for East. These differences between datasets can be attributed to the differences in the underlying NWMs used for NTOL calculation. NTOL is most dominant for coastal points. For LLR stations, this effect is observed for LURE station (not shown).

\subsection{Hydrological loading}\label{subsec:hydl}
HYDL's effect in space geodetic techniques has been studied by \cite{vanDam_etal_2007,tregoning_etal_2009,gfz_ntsl,dill_etal_2018} and others. HYDL is caused by redistribution of continental water mass, such as snow, ground water, et cetera. According to \cite{gfz_ntsl}, it is most dominating along lakes and river sides, and along the Rocky Mountains (North America).

\begin{figure*}[!ht]
	\centering
	\includegraphics[width=0.74\textwidth]{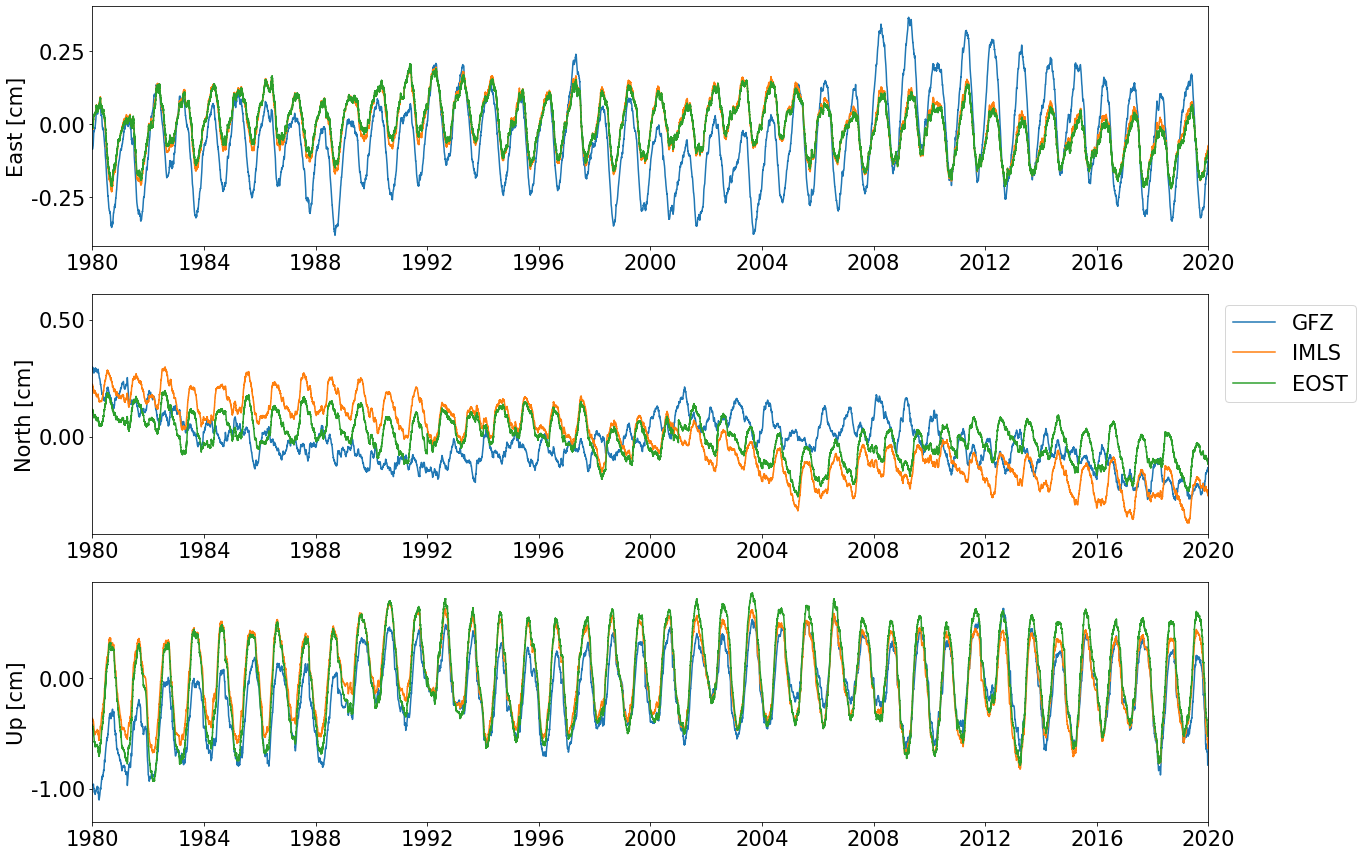}
	\caption{Effect of HYDL at the OCA station from 1980 to 2020 for GFZ, IMLS, and EOST datasets.}\label{fig:OCA_HYDL}
\end{figure*}

Figure \ref{fig:OCA_HYDL} show the HYDL loading at the OCA station. All LLR stations show a similar trend for HYDL loading, i.e. EOST and IMLS HYDL time series are similar (ranging between -0.90 cm and 0.75 cm for Up component, -0.35 cm and 0.30 cm for North component, and -0.20 cm and 0.20 cm for East component), however the time series from the GFZ dataset (ranging between -1.20 cm and 0.60 cm for Up component, -0.25 cm and 0.55 cm for North component, and -0.40 cm and 0.35 cm for East component) differs significantly. The differences between the datasets are assumed to be a resultant of differences between their underlying NWMs and the kinds of water mass redistribution considered therein. As pointed out by \cite{glomsda_etal2020}, LSDM continental water storage (GFZ) considers soil moisture, snow accumulation, seasonal runoff from glaciers, and water flow in river channels given as daily states on a 0.5 degree regular global grid \citep{gfz_ntsl}. MERRA2 (IMLS and EOST), on the other hand, includes snow coverage, soil moisture, stream-flow, and observation-based precipitation, given on on 0.625 degree x 0.5 degree grid \citep{merra2}.


\subsection{Sea level loading}\label{subsec:slel}
In addition to NTAL, NTOL, and HYDL, the GFZ additionally provides another component of NTL, Sea level loading (SLEL), to be considered. Each single loading model conserves its own total mass, but as the GFZ dataset uses different models for NTAL and HYDL, the global mass is not conserved as the mass exchange between the atmosphere and land is not considered \citep{gfz_web}. Therefore, the sea-level varies. Including this change helps obtain a global mass conservation. The SLEL use continental mass from the global hydrological model LSDM, and the atmospheric mass from the model ECMWF. For any point on Earth, the total NTL for GFZ dataset will be the sum of all four loadings. In this study, we consider SLEL as a part of the total NTL at any station for the GFZ dataset, however we do not discuss individual results due to SLEL.

\subsection{Degree 1 deformation}\label{subsec:cm_vs_cf}
Due to the difference in the centre of mass of the Earth to the solid Earth's centre of mass, the loading deformation can be defined in a centre of figure (CF) frame (realised from the positions of geodetic stations on the solid Earth) or centre of mass (CM) frame (centre of orbiting satellites). This is defined by the choice of degree-one load Love numbers, which enter the Green’s function summation \citep{petrov_boy04,gfz_ntsl,imls_ntsl}. For further details on the differences between CM and CF, we refer the reader to \citep{sun_yu_gcm}. The loading obtained from the summation is then defined in the reference frame chosen for the load Love numbers. In our LLR analysis, the a-priori station positions are aligned to CM frame and therefore the loadings from all datasets used within this study were chosen in the CM frame.
\section{Impact of NTL on LUNAR}\label{sec:NTL_LUNAR}
IfE's standard lunar laser ranging analysis software, LUNAR (see, for details \citep{lunar_book} and latest version \citep{hof_mu18}), did not include any non-tidal loading effects so far. Within LUNAR, an adjustment is performed following the Gauss-Markov model, using over 250 parameters, such as the dynamical parameters for the Moon, LLR station and reflector coordinates, et cetera. The post-fit residuals are then obtained after the adjustment. In current version of LUNAR, the a-priori station coordinates are taken from ITRF2014 \citep{itrf2014}, with the exception of coordinates for the APOLLO station which were personally communicated to us by Prof. Thomas Murphy; and the a-priori reflector coordinates are taken from \citep{de421_moon} and \citep{murphy_et_al_2010}. Table \ref{tab:base_res} shows a list of models used within LUNAR, which are all based on the IERS 2010 conventions \citep{Petit2010}.

\begin{table}[!ht]
\caption{List of models used to compute the standard solution.}
  \centering
    \begin{tabular}{p{3.7cm}p{3.7cm}}
    \hline\noalign{\smallskip}
    Effect & Model \\
    \noalign{\smallskip}\hline\noalign{\smallskip}
    Tidal atmospheric delay & Ray and Ponte (2003) \\ \hline
    \rule{0pt}{2.5ex}Tidal ocean loading & FES 2004 \\ \hline
    \rule{0pt}{2.5ex}Solid Earth tides & IERS 2010 \\ \hline
    \rule{0pt}{2.5ex}Deformation due to polar motion & IERS 2010 mean pole model \\ \hline
    \rule{0pt}{2.5ex}Ocean pole tides & Desai (2002) \\ \hline
    \rule{0pt}{2.5ex}Tropospheric delay & Mendes and Pavlis (2004) \\
    \hline
    \end{tabular}%
  \label{tab:base_res}%
\end{table}%

In this study, the effect of NTLs, described in section \ref{sec:ntsl}, was added to LUNAR to analyse the effect on various results. A degree 10 Lagrange interpolation was performed on the time series of all individual loadings in all datasets. The NTL effects were added as corrections to the station coordinates at observation level. To compare if the NTL contribution leads to an improvement or a deterioration, the reference results, hereon be referred to as \enquote*{standard} solution, are compared to the results obtained upon addition of NTL in LUNAR.

The timespan in which the NTL are available, as mentioned in Table \ref{tab:model_list}, are different for the datasets and loadings. For sake of comparison between the results obtained from all datasets, the results shown in this paper are for the case when loadings are added only in the timespan 1980 to present, and loadings with shorter timespans (i.e. NTOL from EOST, NTAL from VMF and ULux) are not added.

In the results, \enquote{NTSL} refers to the solution with all loadings of any dataset applied together in the analysis, i.e. a combination of NTAL, NTOL, HYDL, and SLEL for GFZ; a combination of NTAL, NTOL, and HYDL for IMLS; and a combination of NTAL and NTOL for EOST.

\subsection{One way weighted root mean square of the residuals}
To ascertain if the added effect of NTL in LLR analysis proves to be useful or not, we calculated the change in the one way annually averaged weighted root mean square of the post-fit LLR residuals obtained from LUNAR, henceforth referred to as WRMS.

\begin{figure}[!ht]
    \includegraphics[width=0.49\textwidth]{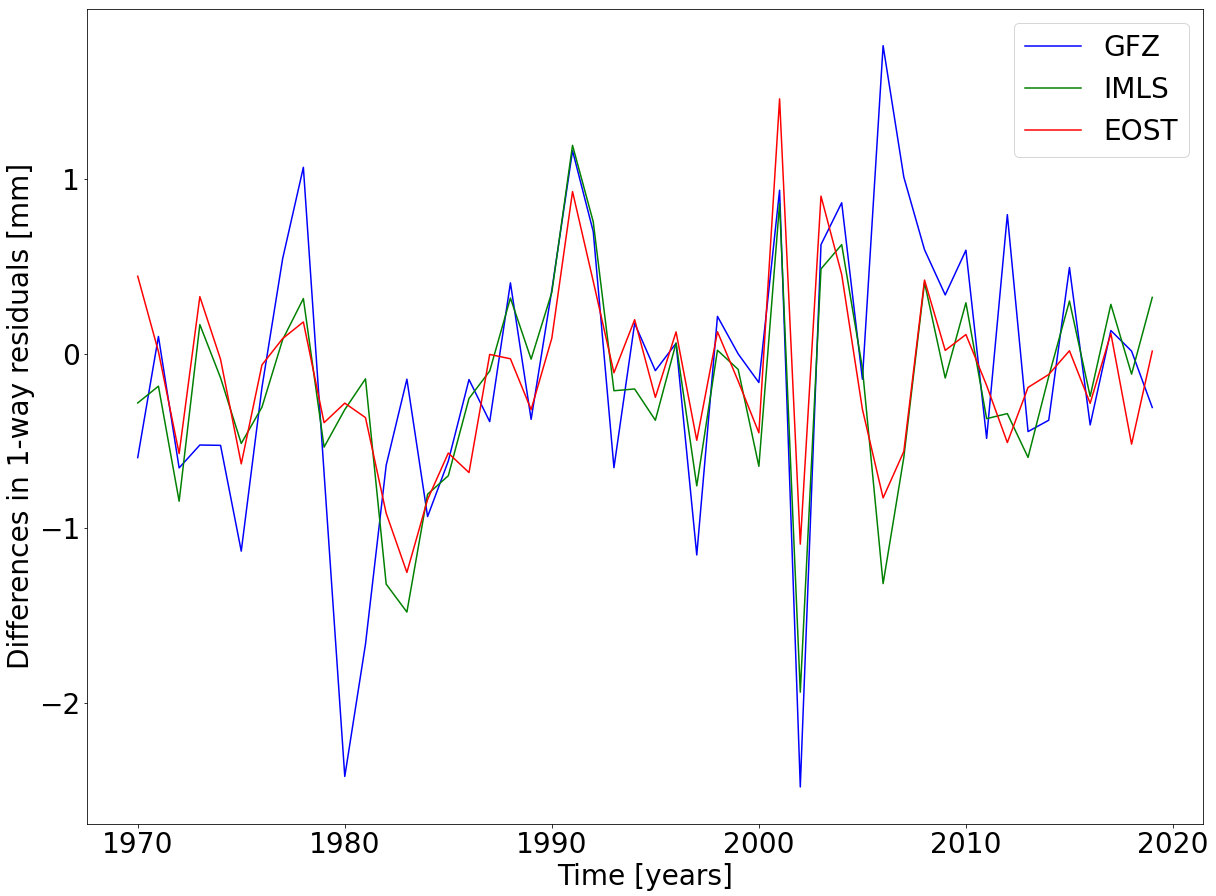}
    \caption{WRMS for GFZ and IMLS  \textbf{NTSL} subtracted from standard solution, for all LLR stations.}\label{fig:wrms_ntsl_diff}
\end{figure}

\begin{figure}[!ht]
    \centering
    \includegraphics[width=0.49\textwidth]{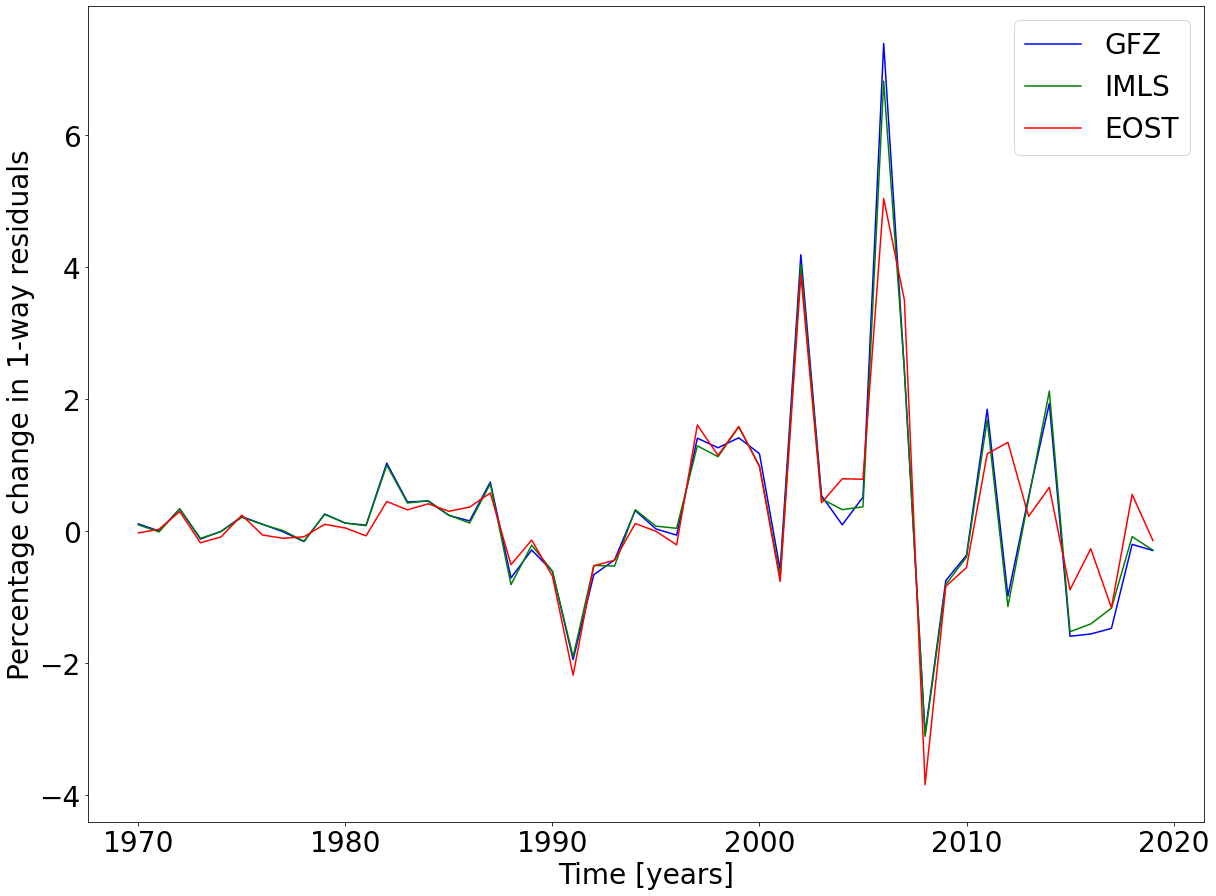}
    \caption{Percentage change in WRMS for the GFZ, IMLS, and EOST \textbf{NTAL} solutions compared to the standard solution for all LLR stations.}\label{fig:wrms_ntal}
\end{figure}

\begin{figure}[!ht]
    \centering
    \includegraphics[width=0.49\textwidth]{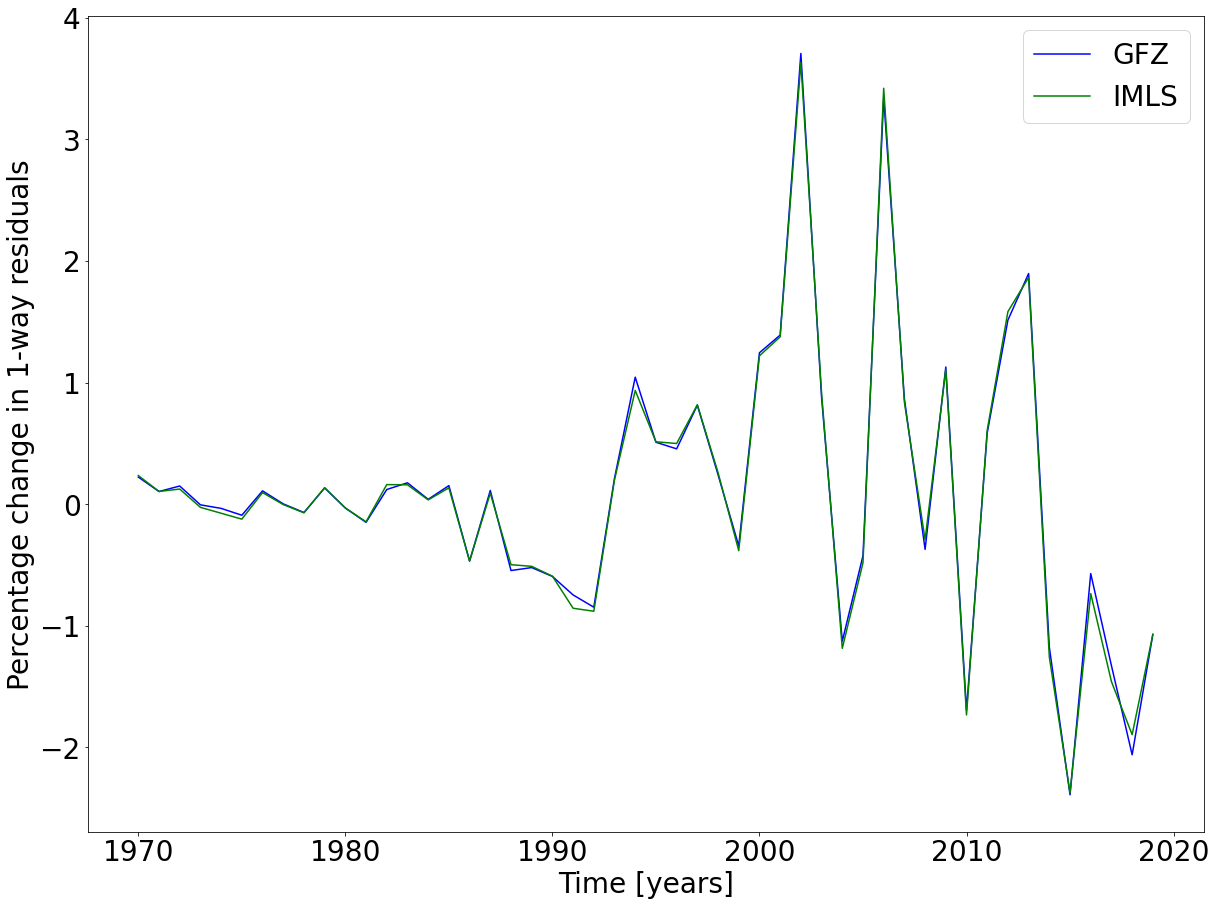}
    \caption{Percentage change in WRMS for the GFZ and IMLS \textbf{NTOL} solutions compared to the standard solution for all LLR stations.}\label{fig:wrms_ntol}
\end{figure}

\begin{figure*}[!ht]
    \begin{minipage}{.48\textwidth}
    \centering
    \includegraphics[width=1\textwidth]{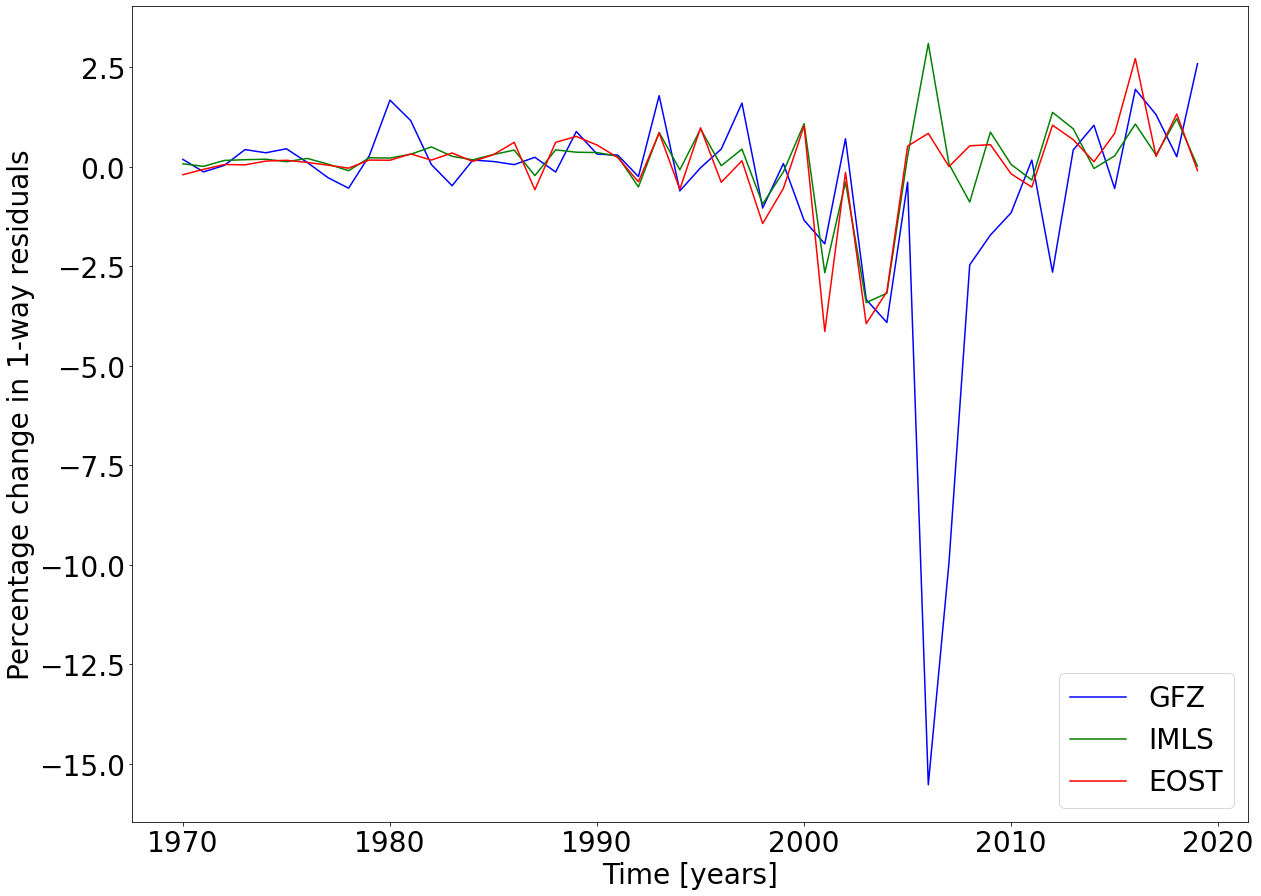}
    \caption{Percentage change in WRMS for the GFZ, IMLS, and EOST \textbf{HYDL} solutions compared to the standard solution for all LLR stations.}\label{fig:wrms_hydl}
    \end{minipage}%
    \qquad
    \begin{minipage}{.48\textwidth}
    \centering
    \includegraphics[width=1\textwidth]{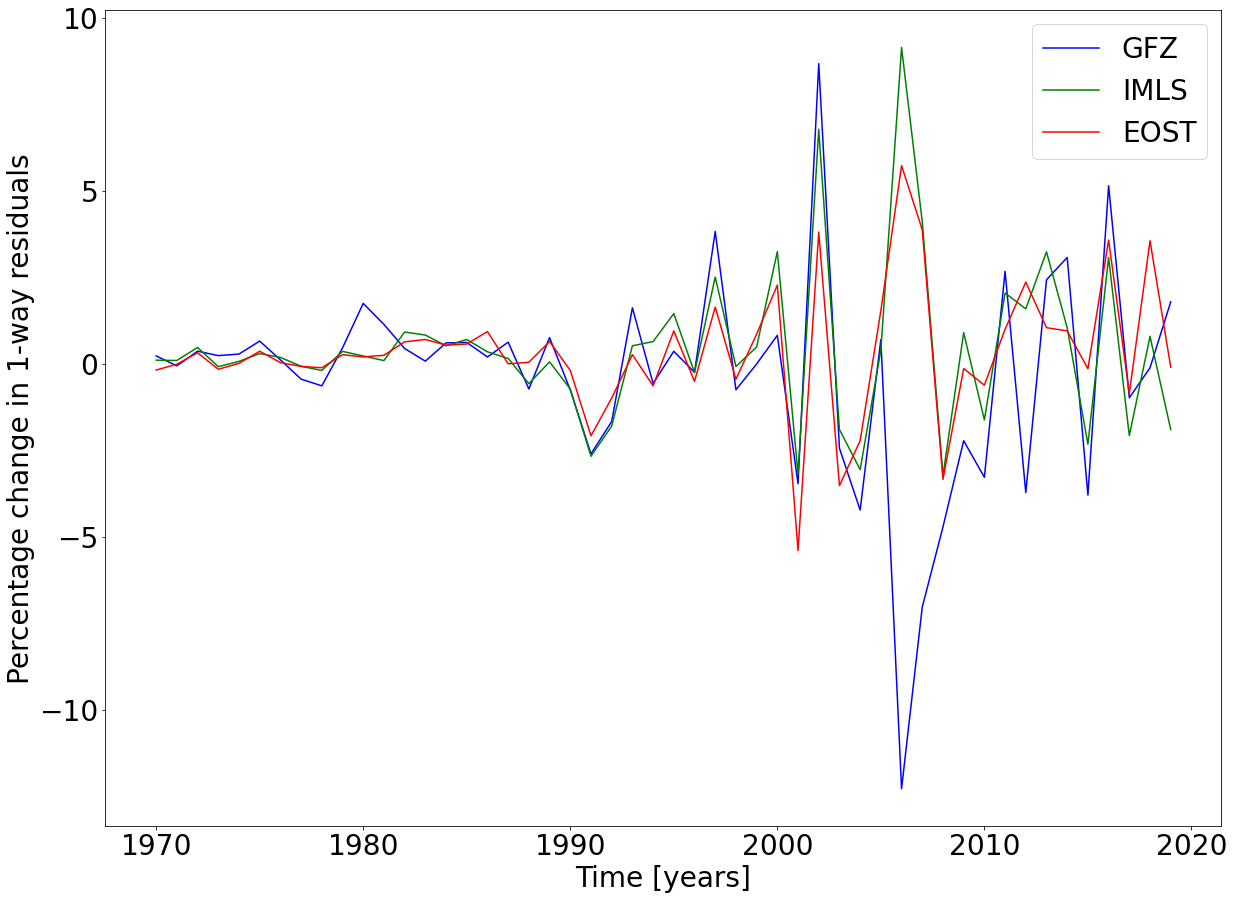}
    \caption{Percentage change in WRMS for the GFZ, IMLS, and EOST \textbf{NTSL} solutions compared to the standard solution for all LLR stations.}\label{fig:wrms_ntsl}
    \end{minipage}
\end{figure*}

Figure \ref{fig:wrms_ntsl_diff} shows the magnitude of differences (standard solution minus NTL solutions) obtained in the WRMS (negative values mean lower residuals, and therefore better), showing the effect impacting up to a few millimetres. Figure \ref{fig:wrms_ntal} to \ref{fig:wrms_ntsl} show the percentage change in WRMS obtained from LUNAR when adding NTL for all LLR stations (positive change means improvement, i.e. lower value of WRMS).

From Figure \ref{fig:wrms_ntal}, it can be seen that the effect of NTAL's addition in LUNAR has similar effects on the post-fit residuals for loadings from all three data centres; the results obtained from IMLS and GFZ datasets almost overlap each other, and EOST is in close agreement. The percentage change due to NTAL from EOST lies between 4\% deterioration and 5\% improvement, and due to NTAL from GFZ and IMLS lies between 3\% deterioration and 7.5\% improvement. A mean value of percentage change per year over all 50 years of data shows an improvement of approximately 0.27\% for all datasets in NTAL. For NTOL (see Figure \ref{fig:wrms_ntol}), the percentage change over the years from IMLS and GFZ datasets overlap each other, ranging between 2.5\% deterioration and 3.7\% improvement, producing an overall improvement in the mean value of percentage change over 50 years of 0.09\%. The percentage change due to HYDL is different for all datasets, lying between 15.5\% deterioration and 2.6\% improvement for GFZ, between 3.5\% deterioration and 3\% improvement for IMLS, and between 4\% deterioration and 3\% improvement for EOST. Overall in the LLR timespan, GFZ shows a deterioration of 0.58\%; however, the HYDL datasets from IMLS and EOST show an improvement in the mean change over the years of 0.10\% and 0.02\%, respectively (see Figure \ref{fig:wrms_hydl}). This difference in the results between the HYDL datasets is expected due to the different input time series, as shown by Figure \ref{fig:OCA_HYDL}.

When combining all loadings from each dataset (see Figure \ref{fig:wrms_ntsl}), the range of change differs for each data set. For GFZ the change ranges between 12\% deterioration and 9\% improvement, for IMLS between 3\% deterioration and 9\% improvement, and for EOST between 5\% deterioration and 6\% improvement. The mean value of percentage change over all years is 0.43\% for EOST (improvement), 0.35\% for IMLS (improvement), and -0.34\% for GFZ (deterioration). Figure \ref{fig:wrms_ntal} to \ref{fig:wrms_ntsl} show a higher value of percentages in the last thirty years because of better laser systems which help obtain a lower value of the LLR residuals in the recent years.

\begin{table*}[!ht]
  \centering
  \caption{Mean values of WRMS for the standard solution (Std), and for the solutions with GFZ, IMLS, and EOST datasets for all LLR stations and loadings.}
    \begin{tabular}{cccccccccc}
    \hline\noalign{\smallskip}
    {\multirow{2}[1]{*}{Observatory}} & \multirow{2}[1]{*}{Dataset} & {GFZ} & {IMLS} & EOST & {\multirow{2}[1]{*}{Observatory}} & \multirow{2}[1]{*}{Dataset} & {GFZ} & {IMLS} & EOST \\
  & {} & {[mm]} & {[mm]} & [mm] &   & {} & {[mm]} & {[mm]} & [mm] \\
    \noalign{\smallskip}\hline\noalign{\smallskip}
    \multirow{6}[1]{*}{APOLLO} & Std & 15.02 & 15.02 & 15.02 & \multirow{6}[1]{*}{LURE} & Std & 64.79 & 64.79 & {64.79} \\
\noalign{\smallskip}
\cline{2-5}\cline{7-10}\noalign{\smallskip}  & NTAL & 14.91 & 14.92 & {14.91} &   & NTAL & 64.67 & 64.72 & {64.41} \\
\noalign{\smallskip}
\cline{2-5}\cline{7-10}\noalign{\smallskip}  & NTOL & 14.95 & 14.95 & - &   & NTOL & 65.11 & 65.10 & - \\
\noalign{\smallskip}
\cline{2-5}\cline{7-10}\noalign{\smallskip}  & HYDL & 15.50 & 14.95 & {14.97} &   & HYDL & 64.57 & 64.72 & {64.54} \\
\noalign{\smallskip}
\cline{2-5}\cline{7-10}\noalign{\smallskip}  & NTSL & 15.42 & 14.77 & 14.84 &   & NTSL & 64.47 & 64.68 & 64.22 \\
    \noalign{\smallskip}\hline\noalign{\smallskip}
    \multirow{6}[1]{*}{McDonald} & Std & 167.77 & 167.77 & {167.77} & \multirow{6}[1]{*}{OCA} & Std & 38.81 & 38.81 & {38.81} \\
\noalign{\smallskip}
\cline{2-5}\cline{7-10}\noalign{\smallskip}  & NTAL & 167.41 & 167.42 & {167.57} &   & NTAL & 38.78 & 38.80 & {38.76} \\
\noalign{\smallskip}
\cline{2-5}\cline{7-10}\noalign{\smallskip}  & NTOL & 167.63 & 167.64 & - &   & NTOL & 38.79 & 38.79 & - \\
\noalign{\smallskip}
\cline{2-5}\cline{7-10}\noalign{\smallskip}  & HYDL & 167.47 & 167.49 & {167.64} &   & HYDL & 38.69 & 38.71 & {38.69} \\
\noalign{\smallskip}
\cline{2-5}\cline{7-10}\noalign{\smallskip}  & NTSL & 167.20 & 167.27 & 167.45 &   & NTSL & 38.60 & 38.67 & 38.64 \\
    \noalign{\smallskip}\hline\noalign{\smallskip}
    \multirow{6}[1]{*}{MLRS1} & Std & 104.98 & 104.98 & {104.98} & \multirow{6}[1]{*}{MLRO} & Std & 31.11 & 31.11 & {31.11} \\
\cline{2-5}\cline{7-10}\noalign{\smallskip}  & NTAL & 104.21 & 104.25 & {104.33} &   & NTAL & 30.88 & 30.86 & {30.82} \\
\noalign{\smallskip}
\cline{2-5}\cline{7-10}\noalign{\smallskip}  & NTOL & 104.93 & 104.95 & - &   & NTOL & 31.42 & 31.40 & - \\
\noalign{\smallskip}
\cline{2-5}\cline{7-10}\noalign{\smallskip}  & HYDL & 104.71 & 105.01 & {105.10} &   & HYDL & 30.71 & 30.91 & {30.79} \\
\noalign{\smallskip}
\cline{2-5}\cline{7-10}\noalign{\smallskip}  & NTSL & 103.99 & 104.39 & 104.46 &   & NTSL & 30.58 & 30.88 & 30.50 \\
    \noalign{\smallskip}\hline\noalign{\smallskip}
    \multirow{6}[1]{*}{MLRS2} & Std & 41.26 & 41.26 & {41.26} & \multirow{6}[1]{*}{WRLS} & Std & 44.19 & 44.19 & {44.19} \\
\noalign{\smallskip}
\cline{2-5}\cline{7-10}\noalign{\smallskip}  & NTAL & 41.27 & 41.26 & {41.28} &   & NTAL & 42.68 & 42.78 & {44.16} \\
\noalign{\smallskip}
\cline{2-5}\cline{7-10}\noalign{\smallskip}  & NTOL & 41.18 & 41.20 & - &   & NTOL & 43.84 & 43.92 & {-} \\
\noalign{\smallskip}
\cline{2-5}\cline{7-10}\noalign{\smallskip}  & HYDL & 41.18 & 40.87 & {40.99} &   & HYDL & 45.08 & 45.06 & {44.11} \\
\noalign{\smallskip}
\cline{2-5}\cline{7-10}\noalign{\smallskip}  & NTSL & 41.24 & 40.89 & 41.02 &   & NTSL & 43.15 & 43.18 & 44.08 \\
    \noalign{\smallskip}\hline
    \end{tabular}%
  \label{tab:wrms_stns}%
\end{table*}%

Table \ref{tab:wrms_stns} shows the mean values of the WRMS obtained for the standard solution, and for solutions with all individual NTL effects, for each station individually. From the table, it can be noticed that HYDL from GFZ shows a deterioration for APOLLO and WLRS stations, but an improvement for the other stations; however, as the magnitude of the deterioration is higher than the magnitude of the improvement, the loading shows an overall deterioration. On the other hand, HYDL from both EOST and IMLS show an improvement in the mean WRMS values of the residuals for all stations except MLRS1 and WLRS. For the the OCA station, which has the highest contribution of NPs (60.73\% of NP data), the performance of HYDL is similar solutions from all three datasets. As the APOLLO station station contributes 9.63\% of NP data, the deterioration of the WRMS there plays a critical role in LLR analysis. The other loadings from all datasets mostly show an improvement in the mean WRMS values of the residuals.

\begin{figure}[!ht]
    \centering
    \includegraphics[width=0.49\textwidth]{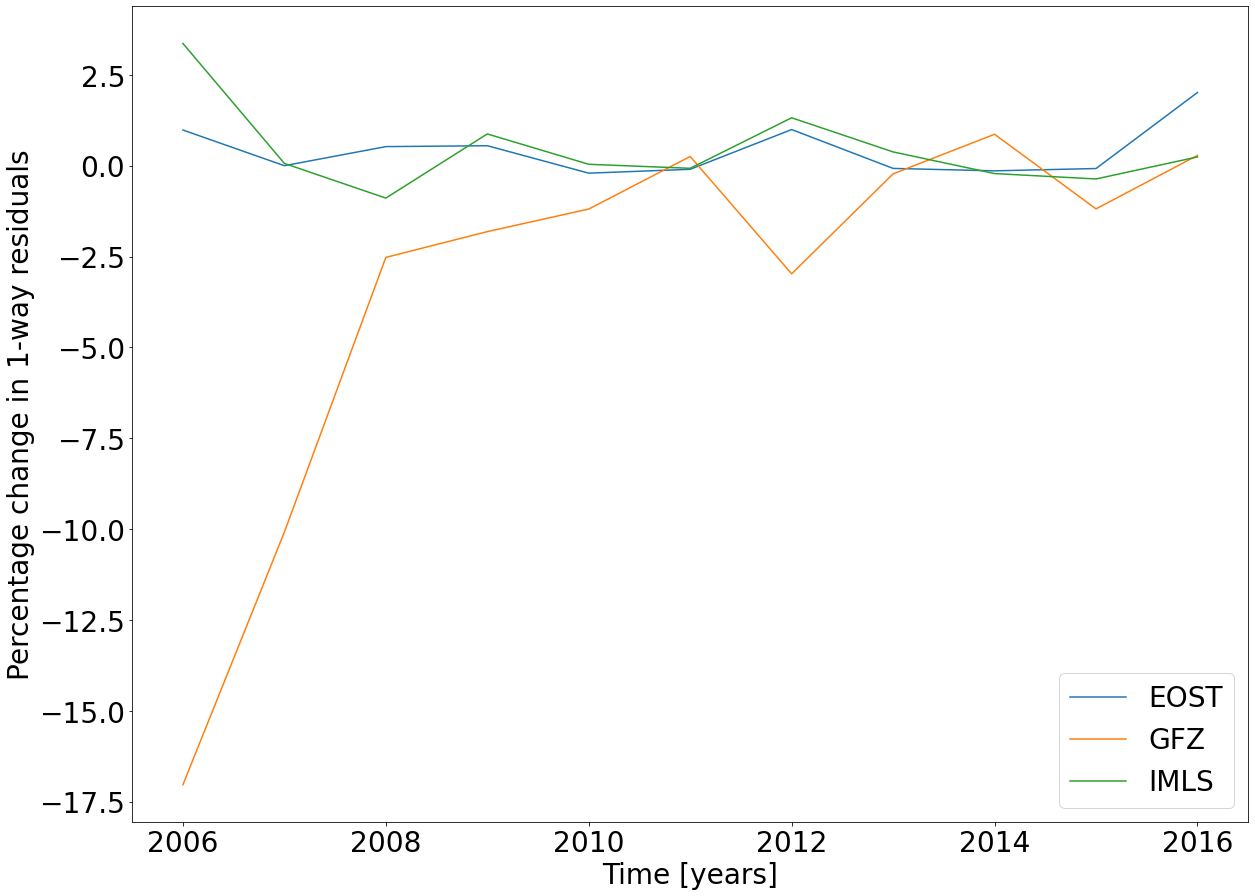}
    \caption{Percentage change in WRMS for the GFZ, IMLS, and EOST \textbf{NTAL} solutions compared to the standard solution for the APOLLO station.}\label{fig:wrms_apollo}
\end{figure}

\begin{figure}[!ht]
    \centering
    \includegraphics[width=0.49\textwidth]{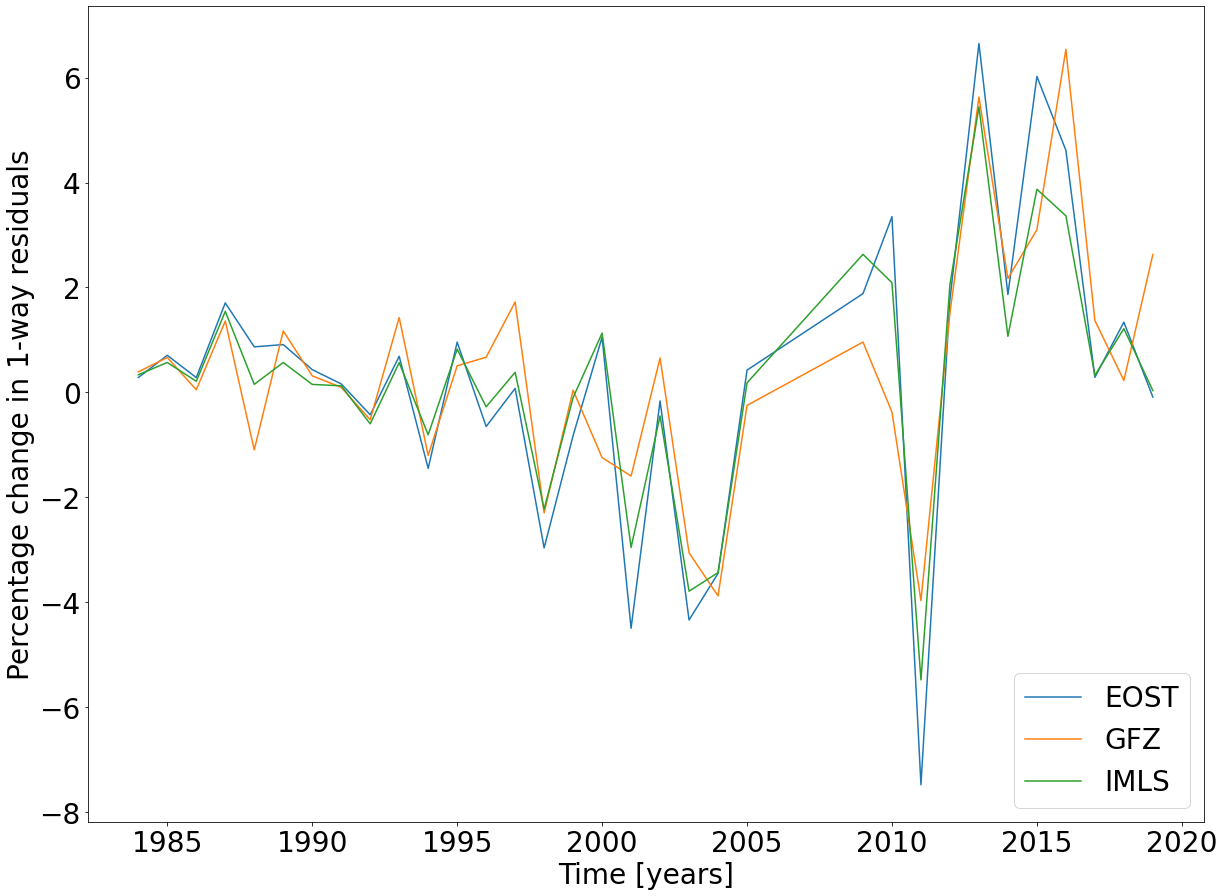}
    \caption{Percentage change in WRMS for the GFZ and IMLS \textbf{NTOL} solutions compared to the standard solution for the OCA station.}\label{fig:wrms_oca}
\end{figure}

Figure \ref{fig:wrms_hydl} shows significant deterioration when adding the GFZ HYDL dataset. However, from Table \ref{tab:wrms_stns}, it can be seen that GFZ HYDL improves the mean WRMS for all stations except APOLLO and WLRS stations. Upon further investigation, it was assessed that the GFZ HYDL leads to a strong deterioration in the WRMS of the annually averaged post-fit residuals only for the APOLLO and McDonald stations (shown only for the APOLLO station as percentage change in WRMS, Figure \ref{fig:wrms_apollo}). In Figure \ref{fig:wrms_apollo}, it can be seen that the percentage change for EOST and IMLS solutions is similar (ranging between -0.20\% and 2\% for EOST, and -1\% and 3.50\% for IMLS), and the GZF solution (ranging between -17\% and 1\%) significantly differs. For the other stations, the solutions from all three datasets show similar percentage change in the WRMS of the annually averaged post-fit residuals. Figure \ref{fig:wrms_oca} shows the percentage change in WRMS for the OCA station, as an example of the similar performance of all three HYDL solutions, as seen by the pattern of the percentage change followed in Figure \ref{fig:wrms_oca}. For the OCA station, the EOST solution ranges between -7.50\% and 6.65\%, the IMLS solution ranges between -5.50\% and 5.50\%, and the GFZ solution ranges between -4\% and 6.50\%.

The GFZ solution shows the most significant deterioration for 2006, as seen in Figure \ref{fig:wrms_hydl} and Figure \ref{fig:wrms_apollo}. This deterioration for 2006, along with a 10\% deterioration for MLRS2 in 2012 (not shown) mainly make the HYDL GFZ and therefore also the NTSL GFZ solutions fall behind the other solutions. The differences between the solutions occur presumably due to the difference in the NWMs between the datasets. The overall agreement with the EOST and IMLS HYDL solutions, and their disagreement with the GFZ HYDL solution indicates that the differences between LSDM and MERRA2 NWMs, mainly over the regions of the APOLLO and McDonald stations, are of importance in LLR analysis. As mentioned in Section \ref{sec:ntsl}, HYDL accounts for mass displacements in the land water, and as the APOLLO and McDonald stations are both surrounded by forest areas of Lincoln National Forest and Davis Mountains State Park, respectively, HYDL and the NWMs used to calculate it most likely play a critical role for these two LLR stations. Overall, it is assessed that adding HYDL from MERRA2 in LUNAR performs better than HYDL from LSDM.

\begin{figure}[!ht]
    \centering
    \includegraphics[width=0.49\textwidth]{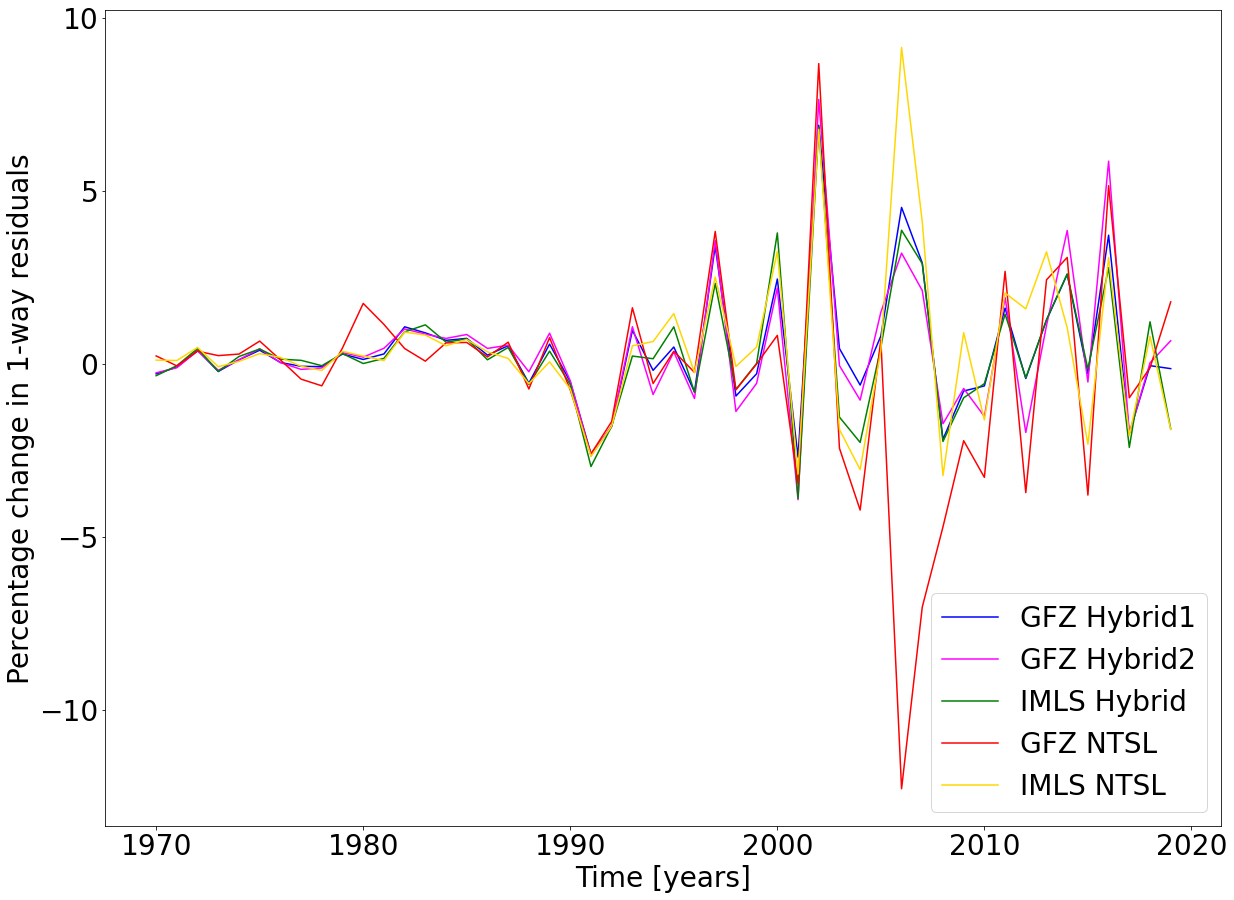}
    \caption{Percentage change in WRMS for the \textbf{hybrid} and \textbf{NTSL} solutions for GFZ and IMLS compared to the standard solution for all stations.}\label{fig:wrms_hybrid}
\end{figure}

To ascertain if the HYDL at the APOLLO and McDonald stations is the only major difference between the datasets, we compute two hybrid solutions of GFZ and IMLS which add all loadings of both datasets (like NTSL) except HYDL and the APOLLO and McDonald stations. For GFZ, two hybrid versions, with and without including SLEL for all LLR stations, are named \enquote{Hybrid1} and \enquote{Hyrbid2}. The percentage change (compared to the standard solution) in the WRMS of the post-fit 1 way LLR residuals of the hybrid and NTSL solutions is shown in  Figure \ref{fig:wrms_hybrid}. It can be seen that the hybrid solutions of GFZ Hybrid1 and IMLS show similar changes in percentages over the years, proving that only the HYDL at the APOLLO and McDonald stations casuses the significant differences between them. The GFZ Hybrid2 performs slightly better, showing that SLEL is an important addition to and a vital aspect of the GFZ datasets. By removing the HYDL at the APOLLO and McDonald stations, the high deterioration in 2006 (at APOLLO, see Figure \ref{fig:wrms_apollo}) is avoided. However, it can also be seen that the IMLS NTSL solution outperforms all three of the hybrid solutions for 2006 and 2012, proving that HYDL at the APOLLO and McDonald stations plays an important role. The mean percentage change over the entire time series for the hybrid solutions are 0.41\%  for GFZ Hybrid1, 0.38\% for GFZ Hybrid2 and 0.25\%  for IMLS.

\subsection{LLR station positions}
In LUNAR, the LLR station coordinates for epoch 2000.0, amongst other parameters, are adjusted. With this adjustment, the uncertainties of the station coordinates are obtained. The mean value of uncertainties (represented as 3-$\sigma$ values) of the coordinates of all six observatories used in LUNAR are given in Table \ref{tab:stn_3sig} for the standard solution as well as for solutions using the NTL datasets. As the McDonald observatory conducted its LLR measurements for different times at three different locations which are very close to each other, namely McDonald, MLRS1, and MLRS2 (linked by local ties), they are analysed as one observatory in LUNAR.

\begin{table*}[!ht]\centering
\caption{Mean values of 3-$\sigma$ uncertainties of LLR station coordinates (estimated for epoch 2000.0) obtained from LUNAR with the standard solution (Std), NTL solutions, and hybrid solutions. }\label{tab:stn_3sig}
\begin{tabular}{ccccccccccc}
\hline\noalign{\smallskip}
\multirow{2}{*}{Observatory} &\multirow{2}{*}{Dataset} &GFZ &IMLS &EOST &\multirow{2}{*}{Observatory} &\multirow{2}{*}{Dataset} &GFZ &IMLS &EOST \\
& &[mm] &[mm] &[mm] & & &[mm] &[mm] &[mm] \\
\noalign{\smallskip}\hline\noalign{\smallskip}
\multirow{7}{*}{APOLLO} &Std &6.65 &6.65 &6.65 &\multirow{7}{*}{LURE} &Std &19.79 &19.79 &19.79 \\
\noalign{\smallskip}
\cline{2-5}\cline{7-10}\noalign{\smallskip} &NTAL &6.66 &6.66 &6.65 & &NTAL &19.82 &19.82 &19.79 \\
\noalign{\smallskip}
\cline{2-5}\cline{7-10}\noalign{\smallskip} &NTOL &6.66 &6.66 &- & &NTOL &19.84 &19.84 &- \\
\noalign{\smallskip}
\cline{2-5}\cline{7-10}\noalign{\smallskip} &HYDL &6.70 &6.63 &6.62 & &HYDL &19.94 &19.73 &19.71 \\
\noalign{\smallskip}
\cline{2-5}\cline{7-10}\noalign{\smallskip} &NTSL &6.71 &6.63 &6.61 & &NTSL &19.97 &19.73 &19.68 \\
\noalign{\smallskip}
\cline{2-5}\cline{7-10}\noalign{\smallskip} &Hybrid1 &6.62 &- &- & &Hybrid1 &19.72 &- &- \\
\noalign{\smallskip}
\cline{2-5}\cline{7-10}\noalign{\smallskip} &Hybrid2 &6.63 &6.63 &- & &Hybrid2 &19.72 &19.75 &- \\
\noalign{\smallskip}\hline\noalign{\smallskip}
\multirow{7}{*}{MLRS2$^{*}$} &Std &9.68 &9.68 &9.68 &\multirow{7}{*}{OCA} &Std &5.38 &5.38 &5.38 \\
\noalign{\smallskip}
\cline{2-5}\cline{7-10}\noalign{\smallskip} &NTAL &9.69 &9.69 &9.68 & &NTAL &5.39 &5.39 &5.38 \\
\noalign{\smallskip}
\cline{2-5}\cline{7-10}\noalign{\smallskip} &NTOL &9.70 &9.70 &- & &NTOL &5.39 &5.39 &- \\
\noalign{\smallskip}
\cline{2-5}\cline{7-10}\noalign{\smallskip} &HYDL &9.75 &9.65 &9.64 & &HYDL &5.42 &5.37 &5.36 \\
\noalign{\smallskip}
\cline{2-5}\cline{7-10}\noalign{\smallskip} &NTSL &9.77 &9.65 &9.62 & &NTSL &5.43 &5.37 &5.35 \\
\noalign{\smallskip}
\cline{2-5}\cline{7-10}\noalign{\smallskip} &Hybrid1 &9.64 &- &- & &Hybrid1 &5.36 &- &- \\
\noalign{\smallskip}
\cline{2-5}\cline{7-10}\noalign{\smallskip} &Hybrid2 &9.65 &9.66 &- & &Hybrid2 &5.36 &5.37 &- \\
\noalign{\smallskip}\hline\noalign{\smallskip}
\multirow{6}{*}{WLRS} &Std &110.42 &110.42 &110.42 &\multirow{6}{*}{MLRO} &Std &29.78 &29.78 &29.78 \\
\noalign{\smallskip}
\cline{2-5}\cline{7-10}\noalign{\smallskip} &NTAL &110.61 &110.60 &110.39 & &NTAL &29.83 &29.83 &29.77 \\
\noalign{\smallskip}
\cline{2-5}\cline{7-10}\noalign{\smallskip} &NTOL &110.68 &110.69 &- & &NTOL &29.85 &29.85 &- \\
\noalign{\smallskip}
\cline{2-5}\cline{7-10}\noalign{\smallskip} &HYDL &111.26 &110.08 &109.99 & &HYDL &30.01 &29.69 &29.66 \\
\noalign{\smallskip}
\cline{2-5}\cline{7-10}\noalign{\smallskip} &NTSL &111.43 &110.11 &109.8 & &NTSL &30.05 &29.70 &29.61 \\
\noalign{\smallskip}
\cline{2-5}\cline{7-10}\noalign{\smallskip} &Hybrid1 &110.03 &- &- & &Hybrid1 &29.67 &- &- \\
\noalign{\smallskip}
\cline{2-5}\cline{7-10}\noalign{\smallskip} &Hybrid2 &110.05 &110.17 &- & &Hybrid2 &29.68 &29.71 &- \\
\noalign{\smallskip}\hline
\end{tabular}
  \begin{flushleft}
    $^{*}$: McDonald, MLRS1, and MLRS2 are linked by local ties and considered as one observatory for adjustment in LUNAR
  \end{flushleft}
\end{table*}

For HYDL, the addition of the GFZ dataset leads to a deterioration for all stations (ranging between 0.72\% and 0.77\%) however, the addition of the IMLS and EOST datasets lead to an improvement for all stations (ranging between 0.19\% and 0.31\% and between 0.37\% and 0.45\%, respectively), also indicating that MERRA2 NWM suits LLR analysis better than LSDM. For NTAL, the addition of the GFZ and IMLS datasets show a slight deterioration (ranging between 0.10\% and 0.19\% from both datasets), whereas the addition of EOST shows either no change or a negligible improvement of up to 0.03\%. For NTOL, addition of both GFZ and IMLS datasets show a deterioration (ranging between 0.15\% and 0.25\% from both datasets). When combining of all loadings from each dataset (represented as \enquote{NTSL} in the table), addition of GFZ shows a deterioration ranging between 0.90\% and 0.93\%, whereas the addition of IMLS and EOST show an improvement ranging between 0.19\% and 0.31\% and between 0.56\% and 0.62\%, respectively.

For the hybrid solutions (without HYDL at the APOLLO and McDonald stations), \enquote{Hyrbid1} and \enquote{Hyrbid2} stand for with and without SLEL for each station, and for IMLS \enquote{Hyrbid2} in the table represents the hybrid solution. The GFZ hybrid solutions show a minor improvement whereas the NTSL solutions show a deterioration, further stressing the importance of HYDL at the APOLLO and McDonald stations. Both the GFZ hybrid solutions perform show very similar results, indicating that SLEL does not have a significant effect on the solutions. For IMLS, the hybrid solution is either similar to, or worse than the NTSL solution, stressing on the importance of HYDL for the APOLLO and McDonald stations and therefore a better suitability of MERRA2 to LUNAR than LSDM.

\subsection{Spectral analysis of LLR residuals}
As the movement of atmospheric, oceanic, and surface water masses is seasonal in nature, it affects the signals obtained from time series of geodetic observations. Many authors, such as \cite{vanDam12,schuh2003,gfz_ntsl} and others, have pointed out the existence of an annual signal in all components of NTL, a semi-annual signal in mainly HYDL, and monthly and half-monthly signals in NTAL and NTOL. The strongest of these signals in all loadings is the annual signal. An addition of NTL in LLR should cause a corresponding effect in the time series of the LLR residuals.

The residuals from LLR contain many different signals, such as monthly, annual, et cetera. The investigation of signals with periods less than one month are difficult with LLR data, as the NPs normally do not cover the span of an entire month due to the lack of LLR observations during new and full Moon, constraining continuity of observations. LLR observations can be further constrained due lower elevations of the Moon, cloudy sky nights, et cetera. In this study, we focus on the annual signal obtained from the LLR time series, which may exist due to different reasons such as unmodelled geocenter motion in LUNAR, affect of asteroids on LLR analysis, et cetera.

The LLR observations are mostly taken at night, with more than one NP per night whenever possible. As the LLR observations can only be taken under certain restrictions (as mentioned above), they are temporally unevenly distributed. In this study, to perform a spectral analysis on this kind of a non-uniformly sampled data, the Lomb-Scargle (LS) periodogram is used. The magnitude of the LLR residuals in LS analysis is not a key factor as the output of the LS periodogram is dimensionless, which is always the case for the standard normalised periodograms.

\begin{figure}[!ht]
    \centering
    \includegraphics[width=0.49\textwidth]{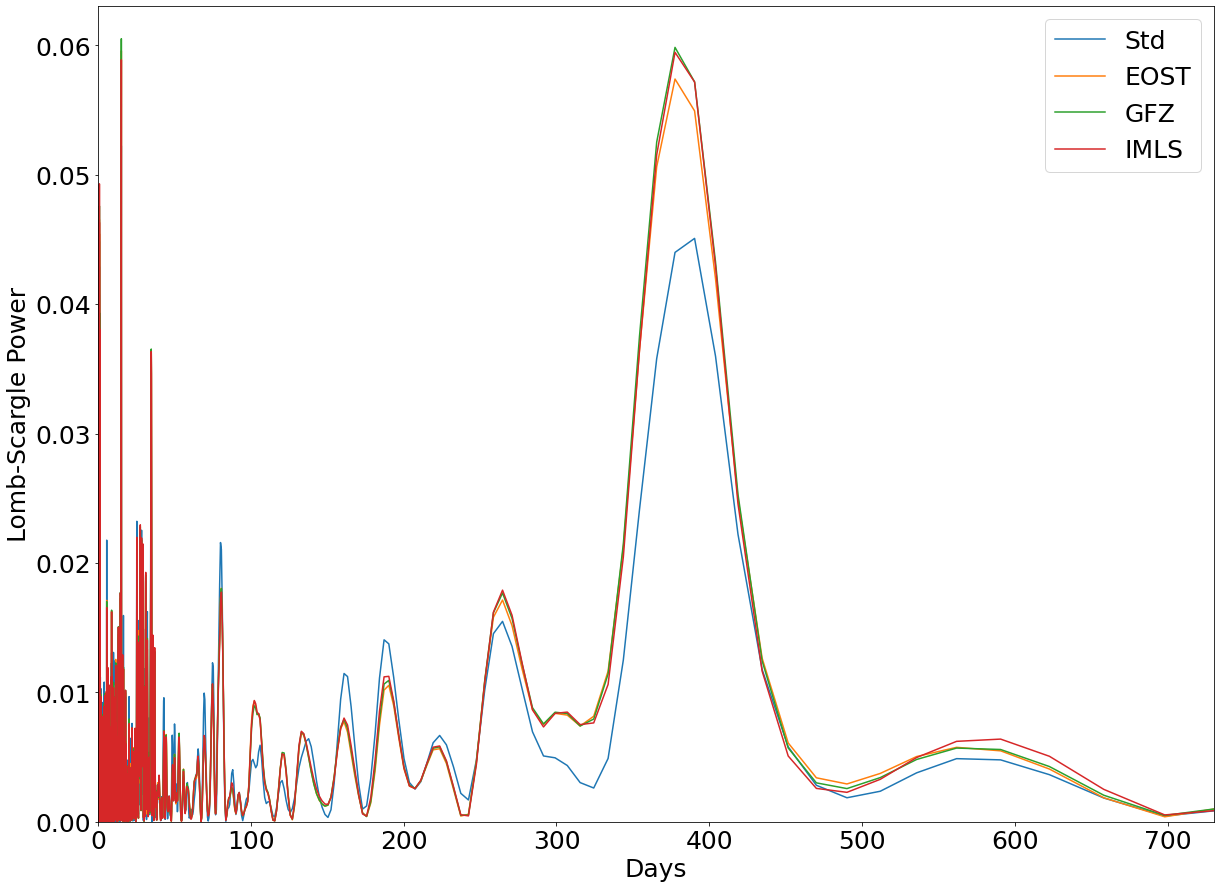}
    \caption{LS periodogram of post-fit LLR residuals obtained at the OCA station from 15.06.2012 to 05.10.2018 for the standard solution and the \textbf{NTAL} solution.}\label{fig:signal_ntal}
\end{figure}

\begin{figure}[!ht]
    \centering
    \includegraphics[width=0.49\textwidth]{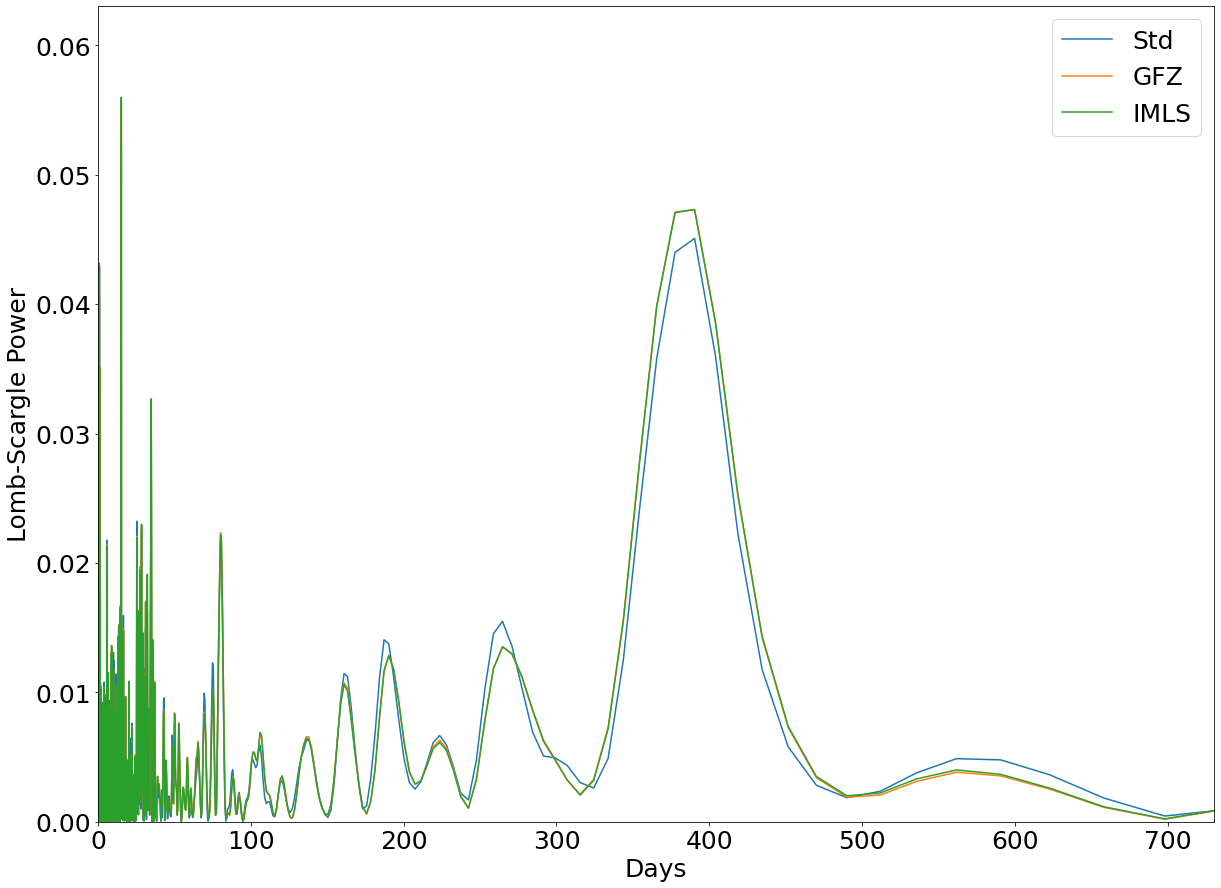}
    \caption{LS periodogram of post-fit LLR residuals obtained at the OCA station from 15.06.2012 to 05.10.2018 for the standard solution and the \textbf{NTOL} solution.}\label{fig:signal_ntol}
\end{figure}

\begin{figure*}[!ht]
    \begin{minipage}{.48\textwidth}
    \centering
    \includegraphics[width=1\textwidth]{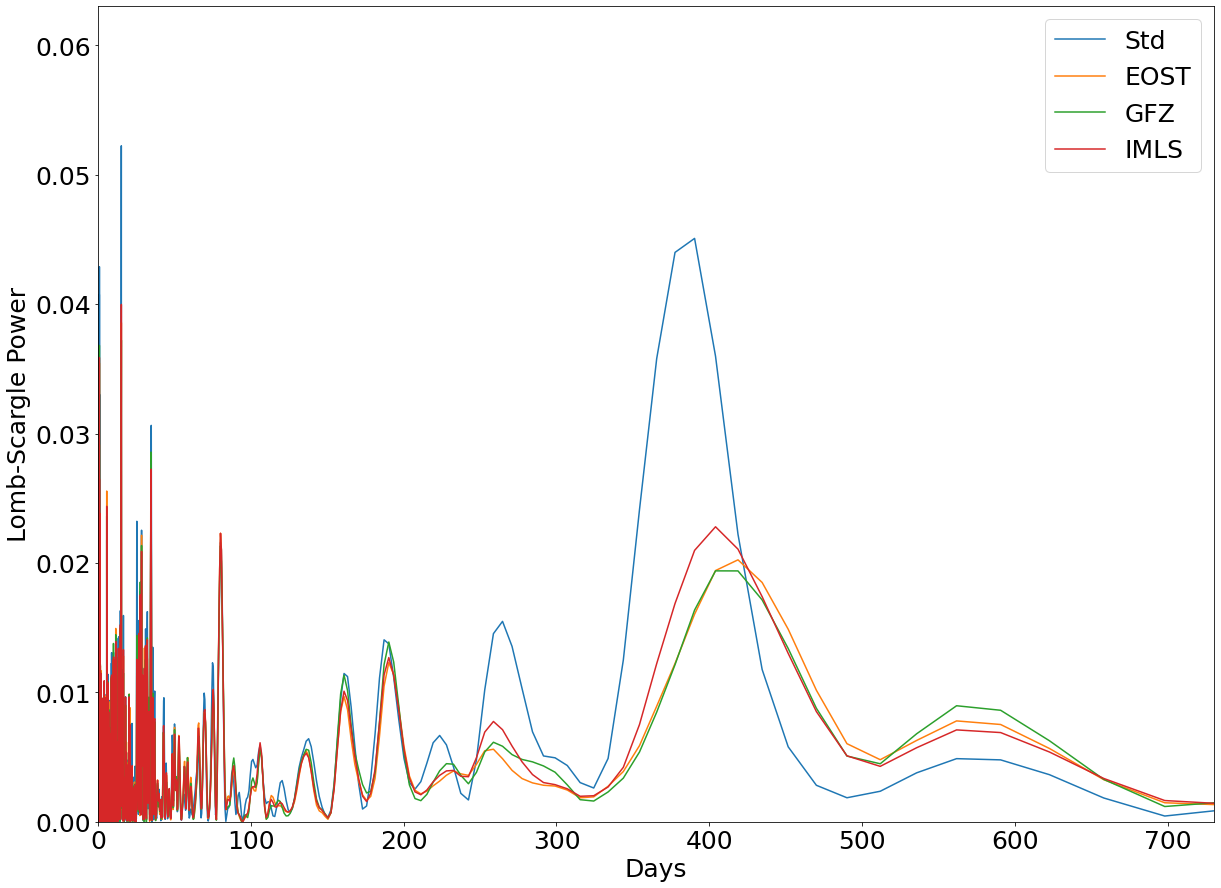}
    \caption{LS periodogram of post-fit LLR residuals obtained at the OCA station from 15.06.2012 to 05.10.2018 for the standard solution and the \textbf{HYDL} solution.}\label{fig:signal_hydl}
    \end{minipage}%
    \qquad
    \begin{minipage}{.48\textwidth}
    \centering
    \includegraphics[width=1\textwidth]{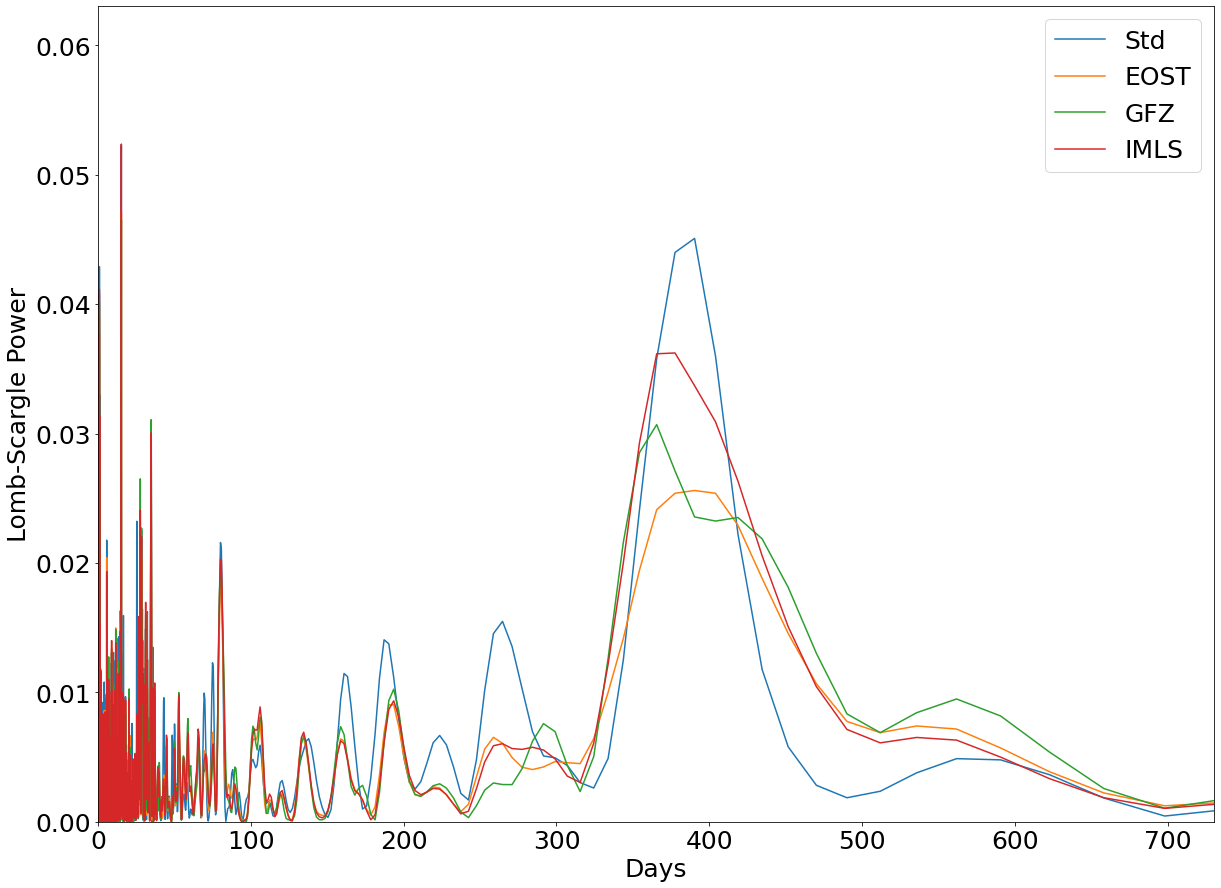}
    \caption{LS periodogram of post-fit LLR residuals obtained at the OCA station from 15.06.2012 to 05.10.2018 for the standard solution and the \textbf{NTSL} solution.}\label{fig:signal_ntsl}
    \end{minipage}
\end{figure*}

To study the annual signal from the post-fit LLR residuals obtained from LUNAR, a suitable subset of the LLR time series (station wise) must be selected. \cite{ls_ana} points out that to obtain a very clear distribution with LS periodogram, a high sampling rate and uniformity of data samples is needed. To best match this criteria, and to get a long enough timespan of residuals at one station, we identify two suitable subsets of time series from the post-fit residuals: from 15.06.2012 to 05.10.2018 at the OCA station (contains 5375 NPs), and 30.06.1994 to 25.01.2003 at MLRS2 (contains 2198 NPs). Figures \ref{fig:signal_ntal} to \ref{fig:signal_ntsl} show the LS periodogram of the post-fit LLR residuals obtained at the OCA station for the standard solution and with solutions upon addition of NTL.


The annual signal observed from the time series is deviates from one year by several days because of the non-uniformity and low sample size of data \citep{Zhang2020}. From Figure \ref{fig:signal_ntal}, it can be seen that when NTAL is added, the power at the annual period increases for all NTOL solutions. Here, all three NTOL solutions, EOST, GFZ, and IMLS, have similar powers, increasing compared to standard solution by 27.27\%, 32.59\%, and 31.93\% respectively. An increase in power of signal at annual period when adding of NTAL is not uncommon, and was also pointed out by \cite{petrov_boy04}, \cite{glomsda_etal2020} and others.

With the addition of NTOL, the power at annual period is not significantly affected, showing an increase of 4.88\% for both solutions in the LS power (see Figure \ref{fig:signal_ntol}), probably because of the small effect of NTOL at the OCA station.

When adding HYDL the power at the annual period decreases significantly for all solutions, as shown by Figure \ref{fig:signal_hydl}. The decrease for EOST solution is of 55.22\%, for GFZ solution of 56.98\%, and for IMLS solution of 49.45\%. \cite{glomsda_etal2020} also observe a decrease in the annual period's power when HYDL is added. As observed at the OCA station, the reduction in power due to addition of HYDL is stronger than the increase in powers due to addition of NTAL and NTOL, individually. Finally, when all loading components for all datasets are added together, shown by Figure \ref{fig:signal_ntsl}, the annual signal shows a reduction in power (for EOST solution of 43.24\%, for GFZ solution of 31.93\%, and for IMLS solution of 19.74\%), presumably due to HDYL's role in the combined loading.

For the hybrid solutions at the OCA station (not shown), as expected, both values of the hybrid solutions are very similar to the NTSL solutions for the two datasets, as the loadings at the OCA station are not affected by the exclusion of HYDL at the APOLLO and McDonald stations.

Similar trends for all individual loadings and for the combined loading are noticed for the annual signal at MLRS2 in the subset time series from 30.06.1994 to 25.01.2003 (not shown). However the power of annual singal for all loadings and also for standard solution observed at MLRS2 is much smaller (LS power of 0.0089 for standard solution), probably because of fewer sample points (NPs) for a signal analysis using the LS periodogram. For the hybrid solutions, as expected, the power at annual signal increases, as the only component of NTL which leads to a reduction in the power at annual signal, i.e., HYDL is not added at McDonald stations.


\section{Conclusions and further scope}\label{sec:conclusions}
In this study, the effect of NTL was applied as observation level corrections in LLR analysis to investigate its effect on the solutions obtained. The NTL was added as three different loading constituents for mass redistribution in atmosphere, oceans, and land water. The effect of NTL within LUNAR are analysed for data from three different data centres: EOST, GFZ, and IMLS due to the long enough time series of loadings available from these centres. Data from other providers is discussed to be in a range similar to the data used within this study. The impact of NTL on LLR analysis was discussed on solutions of WRMS of post-fit one-way LLR residuals, LLR station coordinates, and for the annual signal obtained from the time series of LLR residuals. The overall impact of NTL is determined to be small, however its addition would improve the LLR modelling and would be useful to achieve high accuracy from LLR analysis. Furthermore, NTL will play an important and more significant role when the accuracy of laser signals improves in future.

The impact on WRMS of post-fit one-way LLR residuals from LUNAR using data for each loading from all data providers is similar, except for HYDL which is similar for solutions from EOST and IMLS datasets, but differ for GFZ dataset. GFZ's HYDL at the APOLLO and McDonald stations plays a critical role in deteriorating the results obtained upon the addition of HYDL from GFZ in LUNAR. This is further proved by the implementation of three hybrid solutions, which show the similarity of results if HYDL at the APOLLO and McDonald stations is not considered in solutions using the IMLS and GFZ datasets. Hence proving that for LLR analysis, the NWM MERRA2 leads to better results than LSDM. For the other stations, the results when adding of HYDL from GFZ are similar to addition of HYDL from either EOST or IMLS.

For the uncertainties of LLR station coordinates obtained via a Gauss-Markov adjustment performed within LUNAR, presented in this study as 3-$\sigma$ values, the addition of NTL shows only a small change. GFZ has the maximum influence, showing a deterioration ranging between 0.90\% and 0.93\% for all LLR stations. EOST (with both its loading components) and IMLS (with all three loading components) show an improvement ranging between 0.19\% and 0.31\% and between 0.56\% and 0.62\% for all LLR stations, respectively.

The most significant impact of addition of NTL is observed in the change of the power of the annual period in the post-fit LLR residuals at the OCA station. When HYDL is added, the power at annual period reduces by 55.22\% for EOST, by 56.98\% for GFZ, and by 49.45\% for IMLS. Addition of NTAL and NTOL from all data provider shows an increase in the annual signal's power at the OCA station. A combined solution of all loadings from data providers shows a decrease in the annual signal's power at the OCA station for EOST of 43.24\%, for GFZ of 31.93\%, and for IMLS of 19.74\%.

Based on this study, we conclude that addition of NTL makes a valid contribution in the LLR analysis, as it reduces systematic effects (even if small) which otherwise would smear over to other LLR parameters. The impact of each individual loading from the different data providers is similar, with the exception of HYDL from GFZ. Overall, the addition of NTL in LLR analysis is deemed to be beneficial to achieve smaller LLR residuals and reduced power of the annual signal in the time series of residuals. In a further study, we would discuss the EOP determination from LUNAR, using high accuracy data from OCA station, and estimate the impact of NTL on the EOP.

\section*{Acknowledgements}
We acknowledge with thanks that 50 years of processed LLR data has been obtained under the efforts of the personnel at the Observatoire de la C\^{o}te d'Azur in France, the LURE Observatory in Maui, Hawaii, the McDonald Observatory in Texas, the Apache Point Observatory in New Mexico, the Matera Laser Ranging observatory in Italy, and the Wettzell Laser Ranging System in Germany. This research was funded by the Deutsches Zentrum für Luft- und Raumfahrt (DLR), and Deutsche Forschungsgemeinschaft (DFG, German Research Foundation) under Germany’s Excellence Strategy EXC 2123 QuantumFrontiers, Project-ID 390837967. Further financial supports were from the Strategic Priority Research Program of the Chinese Academy of Sciences (grant nos. XDB23030100 and XDA15017700) and the National Natural Science Foundation of China (project no. 41704013). We would additionally like to thank Franz Hofmann for his contributions to LUNAR, Jean-Paul Boy (University of Strasbourg, France) for providing additional data for this study, and Johannes Böhm (Technical University of Vienna, Austria) for discussions regarding non-tidal loadings.

\section*{Data availability}
LLR data is collected, archived, and distributed under the auspices of the International Laser Ranging Service (ILRS) \citep{pearlman}; and downloaded from the website\footnote{\url{https://cddis.nasa.gov/About/CDDIS_File_Download_Documentation.html}}. The IMLS non-tidal loading dataset is freely available on the IMLS website\footnote{\url{http://massloading.net/}}; and downloaded for the LLR observatories from the pre-computed time series of 1272 space geodesy sites. The GFZ non-tidal loading dataset is freely available on the ESMGFZ Product Repository page\footnote{\url{http://rz-vm115.gfz-potsdam.de:8080/repository/entry/show/Home?entryid=e0fff81f-dcae-469e-8e0a-eb10caf2975b&output=default.html}}; and is downloaded using the script available at the website\footnote{\url{http://rz-vm115.gfz-potsdam.de:8080/repository/entry/show?entryid=362f8705-4b87-48d1-9d86-2cfd1a2b6ac9}}. The VMF non-tidal atmospheric loading dataset is freely available on the VMF website, and downloaded for LLR observatories from the pre-computed time series of SLR stations available at\footnote{\url{https://vmf.geo.tuwien.ac.at/APL_products/SLR/}}. The EOST non-tidal loading dataset is freely available on the EOST website, and downloaded for LLR observatories from the pre-computed time series of 7500 stations available at\footnote{\url{http://loading.u-strasbg.fr/displ_all.php}}. The University of Luxumbourg non-tidal atmospheric loading dataset is freely available on the website\footnote{\url{https://geophy.uni.lu/atmosphere-downloads/}}, and downloaded using a modified version (edited to download loadings for LLR observatories) of the script available at\footnote{\url{https://geophy.uni.lu/uploads/ggfc/atmosphere/ncep/atml_tseries.f}}.

\bibliography{main}

\begin{thebibliography}{}

\bibitem[Altamimi et~al., 2016]{itrf2014}
Altamimi, Z., Rebischung, P., Métivier, L., and Collilieux, X. (2016).
\newblock Itrf2014: A new release of the international terrestrial reference
  frame modeling nonlinear station motions.
\newblock {\em Journal of Geophysical Research: Solid Earth},
  121(8):6109--6131.

\bibitem[Biskupek, 2015]{bisk2015}
Biskupek, L. (2015).
\newblock {\em {Bestimmung der Erdorientierung mit Lunar Laser Ranging}}.
\newblock PhD thesis, Leibniz University Hannover.

\bibitem[Boy and Lyard, 2008]{boy2008_nto}
Boy, J.-P. and Lyard, F. (2008).
\newblock {High-frequency non-tidal ocean loading effects on surface gravity
  measurements}.
\newblock {\em Geophysical Journal International}, 175(1):35--45.

\bibitem[{Bury} et~al., 2019]{bury_2019}
{Bury}, G., {Sośnica}, K., and {Zajdel}, R. (2019).
\newblock Impact of the atmospheric non-tidal pressure loading on global
  geodetic parameters based on satellite laser ranging to gnss.
\newblock {\em IEEE Transactions on Geoscience and Remote Sensing},
  57(6):3574--3590.

\bibitem[Dach et~al., 2010]{dach_etal2010}
Dach, R., Böhm, J., Lutz, S., Steigenberger, P., and Beutler, G. (2010).
\newblock {Evaluation of the impact of atmospheric pressure loading modeling on
  GNSS data analysis}.
\newblock {\em Journal of Geodesy}.

\bibitem[Dill and Dobslaw, 2013]{gfz_ntsl}
Dill, R. and Dobslaw, H. (2013).
\newblock Numerical simulations of global-scale high-resolution hydrological
  crustal deformations.
\newblock {\em JGR: Solid Earth}, 118.
\newblock \url{doi:10.1002/jgrb.50353.}

\bibitem[Dill et~al., 2018]{dill_etal_2018}
Dill, R., Klemann, V., and Dobslaw, H. (2018).
\newblock Relocation of river storage from global hydrological models to
  georeferenced river channels for improved load-induced surface displacements.
\newblock {\em Journal of Geophysical Research: Solid Earth},
  123(8):7151--7164.

\bibitem[Farrell, 1972]{farrell72}
Farrell, W.~E. (1972).
\newblock Deformation of the earth by surface loads.
\newblock {\em Rev. Geophys. and Spac. Phys.}, 10(3):751–797.
\newblock \url{https://doi.org/10.1029/RG010i003p00761}.

\bibitem[Folkner et~al., 2014]{jplde}
Folkner, W.~M., Williams, J.~G., Boggs, D.~H., Park, R.~S., and Kuchynka, P.
  (2014).
\newblock The planetary and lunar ephemerides de430 and de431.
\newblock {\em IPN Progress Report 42-196}.

\bibitem[Gelaro et~al., 2017]{merra2}
Gelaro, R., McCarty, W., Suárez, M.~J., Todling, R., Molod, A., Takacs, L.,
  Randles, C.~A., Darmenov, A., Bosilovich, M.~G., Reichle, R., Wargan, K.,
  Coy, L., Cullather, R., Draper, C., Akella, S., Buchard, V., Conaty, A.,
  da~Silva, A.~M., Gu, W., Kim, G.-K., Koster, R., Lucchesi, R., Merkova, D.,
  Nielsen, J.~E., Partyka, G., Pawson, S., Putman, W., Rienecker, M., Schubert,
  S.~D., Sienkiewicz, M., and Zhao, B. (2017).
\newblock {The Modern-Era Retrospective Analysis for Research and Applications,
  Version 2 (MERRA-2)}.
\newblock {\em Journal of Climate - American Meteorological Society}.

\bibitem[Glomsda et~al., 2020]{glomsda_etal2020}
Glomsda, M., Bloßfeld, M., Seitz, M., and Seitz, F. (2020).
\newblock {Benefits of non-tidal loading applied at distinct levels in VLBI
  analysis}.
\newblock {\em Journal of Geodesy}.

\bibitem[Hersbach et~al., 2018]{era5}
Hersbach, H., de~Rosnay, P., Bell, B., Schepers, D., Simmons, A., Soci, C.,
  Abdalla, S., Balmaseda, M.~A., Balsamo, G., Bechtold, P., Berrisford, P.,
  Bidlot, J., de~Boisséson, E., Bonavita, M., Browne, P., Buizza, R.,
  Dahlgren, P., Dee, D., Dragani, R., Diamantakis, M., Flemming, J., Forbes,
  R., Geer, A., Haiden, T., Hólm, E., Haimberger, L., Hogan, R., Horányi, A.,
  Janisková, M., Laloyaux, P., Lopez, P., Muñoz-Sabater, J., Peubey, C.,
  Radu, R., Richardson, D., Thépaut, J.-N., Vitart, F., Yang, X., Zsótér,
  E., and Zuo, H. (2018).
\newblock Operational global reanalysis: progress, future directions and
  synergies with nwp.
\newblock Technical Report~27, European Centre for Medium Range Weather
  Forecasts, Shinfield Park, Reading, Berkshire RG2 9AX, England.
\newblock \url{https://doi.org/10.21957/tkic6g3wm}.

\bibitem[Hofmann, 2017]{hofmann17}
Hofmann, F. (2017).
\newblock {\em Lunar Laser Ranging –verbesserte Modellierung der
  Monddynamikund Schätzung relativistischer Parameter}.
\newblock PhD thesis, Leibniz University Hannover.

\bibitem[Hofmann et~al., 2018]{hof_etal18}
Hofmann, F., Biskupek, L., and Müller, J. (2018).
\newblock Contributions to reference systems from lunar laser ranging using the
  ife analysis model.
\newblock {\em Journal of Geodesy}, 92:975–987.
\newblock \url{https://doi.org/10.1007/s00190-018-1109-3}.

\bibitem[Hofmann and Müller, 2018]{hof_mu18}
Hofmann, F. and Müller, J. (2018).
\newblock Relativistic tests with lunar laser ranging.
\newblock {\em Classical and Quantum Gravity}, 35.
\newblock \url{doi: https://doi.org/10.1088/1361-6382/aa8f7a}.

\bibitem[Jungclaus et~al., 2013]{mpiom}
Jungclaus, J.~H., Fischer, N., Haak, H., Lohmann, K., Marotzke, J., Matei, D.,
  Mikolajewicz, U., Notz, D., and von Storch, J.~S. (2013).
\newblock {Characteristics of the ocean simulations in the Max Planck Institute
  Ocean Model (MPIOM) the ocean component of the MPI‐Earth system model}.
\newblock {\em Journal of Advances in Modeling Earth Systems}.

\bibitem[K{\"o}nig et~al., 2018]{koenig_etal_2018}
K{\"o}nig, R., Fagiolini, E., Raimondo, J.-C., and Vei, M. (2018).
\newblock A non-tidal atmospheric loading model: On its quality and impacts on
  orbit determination and c20 from slr.
\newblock In Freymueller, J.~T. and S{\'a}nchez, L., editors, {\em
  International Symposium on Earth and Environmental Sciences for Future
  Generations}, pages 189--194, Cham. Springer International Publishing.

\bibitem[Murphy, 2013]{murp13}
Murphy, T.~W. (2013).
\newblock Lunar laser ranging: the millimeter challenge.
\newblock {\em Reports on Progress in Physics}, 76.
\newblock \url{doi:10.1088/0034-4885/76/7/076901}.

\bibitem[Murphy et~al., 2010]{murphy_et_al_2010}
Murphy, T.~W., Adelberger, E., Battat, J., Hoyle, C., Johnson, N., Mcmillan,
  R., Michelsen, E., Stubbs, C., and Swanson, H. (2010).
\newblock Laser ranging to the lost lunokhod~1 reflector.
\newblock {\em Icarus}, 211.

\bibitem[Mémin et~al., 2020]{memim20}
Mémin, A., Boy, J., and Santamaría-Gómez, A. (2020).
\newblock {Correcting GPS measurements for non-tidal loading}.
\newblock {\em GPS Solutions}.

\bibitem[Mémin et~al., 2014]{memin_etal_2014}
Mémin, A., Watson, C., Haigh, I.~D., MacPherson, L., and Tregoning, P. (2014).
\newblock {Non-linear motions of Australian geodetic stations induced by
  non-tidal ocean loading and the passage of tropical cyclones}.
\newblock {\em Journal of Geodesy}.

\bibitem[Müller et~al., 2014]{lunar_book}
Müller, J., Biskupek, L., Hofmann, F., and Mai, E. (2014).
\newblock Lunar laser ranging and relativity.
\newblock In Kopeikin, S., editor, {\em Frontiers in relativistic celestial
  mechanics. Volume 2: Applications and experiments}, volume~2, chapter~3, page
  103–156. Walter de Gruyter, Berlin.
\newblock ISBN: 3110345455, 9783110345452.

\bibitem[Müller et~al., 2009]{mueller2009}
Müller, J., Biskupek, L., Oberst, J., and Schreiber, U. (2009).
\newblock Contribution of lunar laser ranging to realise geodetic reference
  systems.
\newblock In {\em Geodetic Reference Frames. International Association of
  Geodesy Symposia}, volume 134, pages 55--59. Springer Berlin Heidelberg,
  Berlin, Heidelberg.
\newblock \url{https://doi.org/10.1007/978-3-642-00860-3-8}.

\bibitem[Müller et~al., 2012]{mueller_etal12}
Müller, J., Hofmann, F., and Biskupek, L. (2012).
\newblock Testing various facets of the equivalence principle using lunar laser
  ranging.
\newblock {\em Classical and Quantum Gravity - CLASS QUANTUM GRAVITY}, 29.

\bibitem[Müller et~al., 2019]{mueller19}
Müller, J., Murphy, T.~W., Schreiber, U., Shelus, P.~J., Torre, J.~M.,
  Williams, J.~G., Boggs, D.~H., Bouquillon, S., Bourgoin, A., and Hofmann, F.
  (2019).
\newblock Lunar laser ranging: a tool for general relativity, lunar geophysics
  and earth science.
\newblock {\em Journal of Geodesy}, 93:2195–2210.
\newblock \url{doi: https://doi.org/10.1007/s00190-019-01296-0}.

\bibitem[Nordman et~al., 2015]{nordman_etal2015}
Nordman, M., Virtanen, H., Nyberg, S., and Mäkinen, J. (2015).
\newblock {Non-tidal loading by the Baltic Sea: Comparison of modelled
  deformation with GNSS time series}.
\newblock {\em GeoResJ}.

\bibitem[Oreiro et~al., 2018]{orerio_etal_2018}
Oreiro, F.~A., Wziontek, H., Fiore, M. M.~E., D’Onofrio, E.~E., and Brunini,
  C. (2018).
\newblock {Non-Tidal Ocean Loading Correction for the Argentinean-German
  Geodetic Observatory Using an Empirical Model of Storm Surge for the Río de
  la Plata}.
\newblock {\em Pure and Applied Geophysics}.

\bibitem[{Otsubo} et~al., 2004]{otsubo_04}
{Otsubo}, T., {Kubo-Oka}, T., {Gotoh}, T., and {Ichikawa}, R. (2004).
\newblock {Atmospheric Loading Blue-Sky Effects on SLR Station Coordinates}.
\newblock In {\em AGU Fall Meeting Abstracts}, volume 2004, pages G31B--0793.

\bibitem[Pavlov et~al., 2016]{pav16}
Pavlov, D.~A., Williams, J.~G., and Suvorkin, V.~V. (2016).
\newblock Determining parameters of moon’s orbital and rotational motion from
  llr observations using grail and iers-recommended models.
\newblock {\em Celest Mech Dyn Astr}, 126:61--88.
\newblock \url{https://doi.org/10.1007/s10569-016-9712-1}.

\bibitem[Pearlman et~al., 2002]{pearlman}
Pearlman, M.~R., Degnan, J.~J., and Bosworth, J.~M. (2002).
\newblock The international laser ranging service.
\newblock {\em Advances in Space Research}.
\newblock \url{https://doi.org/10.1016/S0273-1177(02)00277-6}.

\bibitem[Petit and Luzum, 2010]{Petit2010}
Petit, G. and Luzum, B., editors (2010).
\newblock {\em {IERS Conventions 2010}}.
\newblock Number~36 in IERS Technical Note. Verlag des Bundesamtes f\"ur
  Kartographie und Geod\"asie, Frankfurt am Main.

\bibitem[Petrov, 2015]{imls_ntsl}
Petrov, L. (2015).
\newblock The international mass loading service.
\newblock \url{http://arxiv.org/abs/1503.00191}.

\bibitem[Petrov and Boy, 2004]{petrov_boy04}
Petrov, L. and Boy, J.~P. (2004).
\newblock Study of the atmospheric pressure loading signal in very long
  baseline interferometry observations.
\newblock {\em JGR: Solid Earth}.
\newblock \url{https://doi.org/10.1029/2003JB002500}.

\bibitem[Schuh et~al., 2004]{schuh2003}
Schuh, H., Estermann, G., Cr{\'e}taux, J.-F., Berg{\'e}-Nguyen, M., and van
  Dam, T. (2004).
\newblock Investigation of hydrological and atmospheric loading by space
  geodetic techniques.
\newblock In Hwang, C., Shum, C.~K., and Li, J., editors, {\em Satellite
  Altimetry for Geodesy, Geophysics and Oceanography}, pages 123--132, Berlin,
  Heidelberg. Springer Berlin Heidelberg.

\bibitem[Sośnica et~al., 2013]{sonsica_13}
Sośnica, K., Thaller, D., Dach, R., Jäggi, A., and G., B. (2013).
\newblock {Impact of loading displacements on SLR-derived parameters and on the
  consistency between GNSS and SLR results}.
\newblock {\em Journal of Geodesy}.

\bibitem[Sun, 2017]{sun_yu_gcm}
Sun, Y. (2017).
\newblock {\em {Estimating geocenter motion and changes in the Earth’s
  dynamic oblateness from GRACE and geophysical models}}.
\newblock PhD thesis, TU Delft Physical and Space Geodesy.
\newblock \url{10.4233/uuid:7fe64dde-7fb5-4392-8160-da6f7916dc6b}.

\bibitem[Thomas et~al., 2020]{gfz_web}
Thomas, M., Dill, R., and Dobslaw, H. (2020).
\newblock Sea-level loading product description.
\newblock
  \url{http://rz-vm115.gfz-potsdam.de:8080/repository/entry/show?entryid=0612018a-3ba4-44bc-86d9-1a429749fe4d},
  Last check: 18.05.2020.

\bibitem[Tregoning and Watson, 2009]{tregoning_2009}
Tregoning, P. and Watson, C. (2009).
\newblock {Atmospheric effects and spurious signals in GPS analyses}.
\newblock {\em JGR: Solid Earth}.

\bibitem[Tregoning and Watson, 2011]{tregoning_2011}
Tregoning, P. and Watson, C. (2011).
\newblock {Correction to “Atmospheric effects and spurious signals in GPS
  analyses"}.
\newblock {\em JGR: Solid Earth}.

\bibitem[Tregoning et~al., 2009]{tregoning_etal_2009}
Tregoning, P., Watson, C., Ramillien, G., McQueen, H., and Zhang, J. (2009).
\newblock {Detecting hydrologic deformation using GRACE and GPS}.
\newblock {\em Geophysical Research Letters}, 36(15).

\bibitem[van Dam et~al., 1994]{vanDam_etal_1994}
van Dam, T., Blewitt, G., and Heflin, M. (1994).
\newblock {Atmospheric pressure loading effects on Global Positioning System
  coordinate determinations}.
\newblock {\em JGR: Solid Earth}.

\bibitem[van Dam et~al., 2012]{vanDam12}
van Dam, T., Collilieux, X., Wuite, J., Altamimi, Z., and Ray, J. (2012).
\newblock Nontidal ocean loading: amplitudes and potential effects in gps
  height time series.
\newblock {\em Journal of Geodesy}, 86:1043–1057.
\newblock \url{doi: 10.1007/s00190-012-0564-5}.

\bibitem[Van~Dam et~al., 2007]{vanDam_etal_2007}
Van~Dam, T., JM, W., and Lavallée, D. (2007).
\newblock A comparison of annual vertical crustal displacements from gps and
  gravity recovery and climate experiment (grace) over europe.
\newblock {\em Journal of Geophysical Research}, 112.

\bibitem[Van~Dam and Wahr, 1987]{vanDam_Wahr_1987}
Van~Dam, T.~M. and Wahr, J.~M. (1987).
\newblock Displacements of the earth's surface due to atmospheric loading:
  Effects on gravity and baseline measurements.
\newblock {\em Journal of Geophysical Research: Solid Earth},
  92(B2):1281--1286.

\bibitem[VanderPlas, 2017]{ls_ana}
VanderPlas, J.~T. (2017).
\newblock Understanding the lomb-scargle periodogram.
\newblock {\em The Astrophysical Journal Supplement Series}, 236.
\newblock \url{doi: 10.3847/1538-4365/aab766}.

\bibitem[Viswanathan et~al., 2019]{viswa19}
Viswanathan, V., Rambaux, N., andJ. Laskar, A.~F., and Gastineau, M. (2019).
\newblock Observational constraint on the radius and oblateness of the lunar
  core-mantle boundary.
\newblock {\em Geophysical Research Letters}, 46:7295–7303.
\newblock \url{doi: https://doi.org/10.1029/2019GL082677}.

\bibitem[Williams, 2008]{de421_moon}
Williams, J.~G. (2008).
\newblock De421 lunar orbit, physical librations, and surface coordinates.
\newblock {\em Inteoffice Memorandum}.

\bibitem[Williams et~al., 2013]{de430_other}
Williams, J.~G., Boggs, D.~H., and Folkner, W.~M. (2013).
\newblock {DE430 Lunar Orbit, Physical Librations, and Surface Coordinates}.
\newblock Technical report, Jet Propulsion Laboratory, California Institute of
  Technology, Pasadena, CA, USA.
\newblock Interoffice memorandum, IOM 335-JW,DB,WF-20130722-016.

\bibitem[Williams et~al., 2006]{will06}
Williams, J.~G., Turyshev, S.~G., Boggs, D.~H., and Ratcliff, J.~T. (2006).
\newblock Lunar laser ranging science: Gravitational physics and lunar interior
  and geodesy.
\newblock {\em Advances in Space Research}, 37:67--71.
\newblock \url{doi: https://doi.org/10.1016/j.asr.2005.05.013}.

\bibitem[Williams and Penna, 2011]{williams_and_penna}
Williams, S. D.~P. and Penna, N.~T. (2011).
\newblock {Non‐tidal ocean loading effects on geodetic GPS heights}.
\newblock {\em JGR: Solid Earth}.

\bibitem[Wunsch and Stammer, 1997]{wunsch_stammer_97}
Wunsch, C. and Stammer, D. (1997).
\newblock {Atmospheric loading and the oceanic “inverted barometer”
  effect}.
\newblock {\em Reviews of Geophysics}.

\bibitem[Zhang et~al., 2020]{Zhang2020}
Zhang, M., M{\"{u}}ller, J., and Biskupek, L. (2020).
\newblock {Test of the equivalence principle for galaxy's dark matter by lunar
  laser ranging}.
\newblock {\em Celestial Mechanics and Dynamical Astronomy}, 132(4):25.

\end{thebibliography}

\end{document}